\documentclass[11pt]{article}

\usepackage[T1]{fontenc}
\usepackage[utf8]{inputenc}
\usepackage[margin=1in]{geometry}
\usepackage{newtxtext,newtxmath}
\usepackage{microtype}
\usepackage{amsmath}
\usepackage{graphicx}
\usepackage{booktabs,longtable,array}
\usepackage[section]{placeins}
\usepackage{xcolor}
\usepackage[round,authoryear]{natbib}
\usepackage[colorlinks=true,linkcolor=blue!50!black,citecolor=blue!50!black,urlcolor=blue!50!black]{hyperref}
\hypersetup{pdftitle={Decision and Support Mechanisms under Platform Information Overload: A Generative-Agent Experiment on College Application}}

\graphicspath{{figures/}}
\title{\textbf{Decision and Support Mechanisms under Platform Information Overload}\\[9pt]
{\mdseries\fontsize{13}{16}\selectfont A Generative-Agent Experiment on College Application}}

\author{
  \textbf{Tingyu Dai}\textsuperscript{*}\\
  {\small Sociology, Columbia University}\\
  {\small \href{mailto:td2853@columbia.edu}{\texttt{td2853@columbia.edu}}}
  \and
  \textbf{Xuanyou Liu}\textsuperscript{*}\\
  {\small Computer Science, Northwestern University}\\
  {\small \href{mailto:xuanyouliu2030@u.northwestern.edu}{\texttt{xuanyouliu2030@u.northwestern.edu}}}
}
\date{\small \textsuperscript{*}Equal contribution.}

\begin{document}
\maketitle

\begin{abstract}
\noindent
People now make major personal decisions surrounded by far more online information than they can read. This abundance does not necessarily lead to better decisions: research on information overload shows that, because attention is limited, more information can worsen decisions. That research says less about why overload affects some people more than others. We study this question using college applications in China, a high-stakes decision made amid heavy online information, where we can objectively measure the quality of each decision once the admission cutoffs are released.

We simulated 90 applicants using language-model agents, with family backgrounds taken from a national household survey. Each applicant wrote a list of eight programs alone and then revised it four times: after advice from a parent and a teacher, after reading a feed of twenty real online items about applications, after a mentor gave advice based on the applicant's own rank, and after a second mentor offered only emotional support. After each step, we scored the list and asked how anxious the applicant felt. We asked how family socioeconomic background affects who benefits from information, and whether personalized advice or emotional support improves decisions.

When applicants wrote their lists alone, before family background entered the decision, admission rates were similar across SES groups. The gap appeared once family background entered through the parent and teacher. Low-SES parents asked for low tuition and schools close to home, which left their children with fewer programs they could get into. Admission rose to 100\% among high-SES applicants and fell to 77\% among low-SES applicants. The platform information feed gave everyone the same amount of information and could have narrowed this gap. However, it did not. Applicants maintained their decisions but reported higher anxiety. Personalized advice cut the gap to 3 points by raising low-SES admission from 80\% to 97\%. However, it led students to schools with higher tuition and majors they are less interested in. Emotional support did not further lower anxiety, but helped applicants go back and use helpful information items they had ignored earlier.

These early simulation results suggest that when information is abundant, more of it neither improves decisions nor reduces inequality. Personalized information does both, but it can move inequality from the outcome to its cost.

\medskip
\noindent\textbf{Keywords:} platform information overload; decision-making; educational inequality; generative agents
\end{abstract}

\section{Introduction}\label{introduction}

Not long ago, information overload was mostly a problem for executives and analysts. Today it is part of everyone's life. News apps, short videos, influencer channels, ranking sites, and AI assistants surround most big personal decisions, such as choosing a medical treatment, a pension plan, or school applications. Research on information overload has captured this trend and extended it to social media and other new settings \citep{roetzel2019information}. One hopeful view is that this abundance makes things fairer: people who once made decisions based on limited and biased information can now read much of what insiders know. However, much research points the other way. When news flow on a topic increases, people with more education learn more from it than those with less, so the knowledge gap can widen \citep{tichenor1970mass}. The same pattern appeared with the Internet: as access spread, the divide moved from who is online to who has the skills to use it and what they use it for \citep{hargittai2002second,vandeursen2014digital}. Abundance also has its own costs. Once people have more information than they can process, more of it stops improving their decisions and starts to cost them time, attention, and peace of mind \citep{eppler2004concept}.

This literature shows who gains knowledge from abundant information and what it costs. It leaves two things open: (1) whether, in a high-stakes decision, that information narrows or widens the gap in decision quality between people of different socioeconomic status (SES), and (2) what kind of help can close that gap, and at what cost. The closest evidence comes from studies of educational choice, where information programs have been tested at scale, and the results are mixed. Some information helps. In Chile, personalized information about admission chances reduced application mistakes at national scale \citep{larroucau2025college}. In China, information about the earnings associated with different majors changed the choices of disadvantaged applicants \citep{ding2021information}. General information often does not help. In the H\&R Block experiment, giving families information about financial aid without helping them fill in the form had no effect \citep{bettinger2012role}. Nudge campaigns that worked locally had no measurable effect when extended to hundreds of thousands of students \citep{bird2021nudging}. Showing students past admission outcomes can even backfire \citep{tomkins2023showing}.

These mixed results call for an experiment that gives the same people general information and then personal advice, scores their decisions after each, and compares SES groups along the way. Such an experiment is hard to run with real people. We cannot observe the same person with and without a feed, most personal decisions cannot be scored objectively afterward, and the emotional cost of information is rarely measured in most relevant surveys. We therefore run it as a simulation with generative agents, which are simulated people driven by a large language model \citep{park2023generative,piao2025agentsociety}. With agents, we can control exactly what each decision maker sees, interview them after every step, and score their decisions against outcomes they cannot see. Agents also have known weaknesses: they vary less than real people do and can flatten differences between groups \citep{bisbee2024synthetic,wang2025large}. We therefore use them to study mechanisms, not to stand in for real people.

\begin{figure}[!htbp]
\centering
\includegraphics[width=\textwidth]{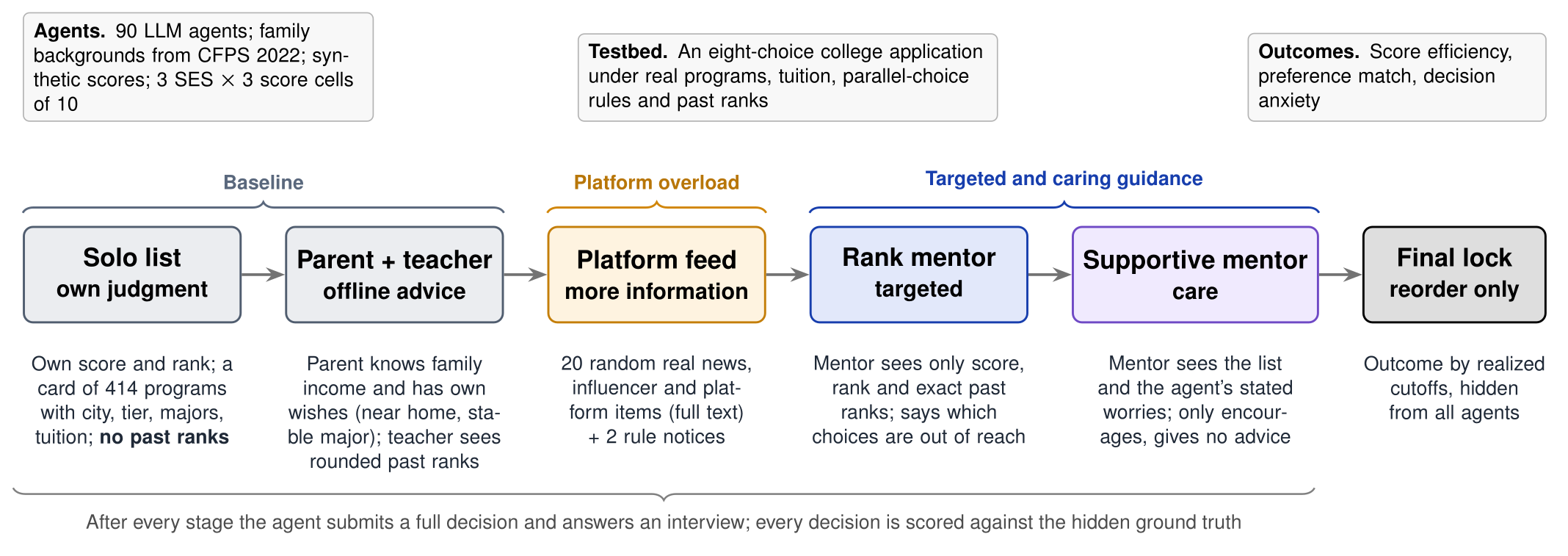}
\caption{Design of the experiment. The same agent goes through every stage in order. The platform feed adds a large volume of information. The two mentors add information aimed at the agent and emotional support. The rank-mentor stage starts from the exact state each agent reached after the feed, and the supportive mentor continues from the state after the rank mentor. The realized cutoffs (the ground truth) are used only for scoring.}
\label{fig:1}
\end{figure}

The setting is college applications in China. It is a high-stakes decision surrounded by platform content: after one influencer commented on journalism, admission ranks for that major shifted measurably \citep{fan2025public}, and families use AI tools that draft application lists \citep{chen2025scores}. It can also be scored objectively by comparing applicants' school lists with the final admission score cutoff. We built the experiment on the AgentSociety\textsuperscript{2} platform \citep{piao2026agentsociety2}. Ninety agents, with different family backgrounds from the China Family Panel Studies \citep{xie2014introduction}, each rank eight programs from a real provincial admission market under the parallel-choice rule \citep{chen2017chinese}. Each agent writes a list alone and then revises it four times, and each revision adds one kind of information (Figure~\ref{fig:1}). Advice from a parent and a teacher brings in the family's circumstances. A feed of twenty real online items represents abundant general information. A mentor who knows the agent's rank (the rank mentor) gives short personal advice. A second mentor (the supportive mentor) offers only encouragement, which separates emotional support from information. After every stage, we score the list on \emph{score efficiency} (how well it uses the agent's exam score) and \emph{preference match} (how closely it fits what the agent wants), and we ask the agent to rate their \emph{decision anxiety} on a scale from 1 to 7. The three measures are defined in Section~\ref{study-design}.

When agents made decisions without SES emphasized, admission rates were similar. The gap appeared once the different SES backgrounds were injected with parent and teacher: low-SES parents asked for low tuition and schools close to home, and low-SES agents were admitted 23 percentage points less often than high-SES agents. The platform information feed did not narrow this gap. The feed was four times as long as the parent and teacher advice, but only about a tenth of it was relevant to the agent and used. Decisions did not improve with more information, but agents' decision anxiety rose, and the gap stayed at 20 points. The application mentor's personal advice nearly closed the SES gap. Admission rose from 90\% to 99\% overall and from 80\% to 97\% among low-SES agents, and the gap fell to 3 points. The cost was that agents gave up majors they were interested in, and low-SES agents were steered toward more expensive and more distant programs. The emotionally supportive mentor did not lower anxiety further, but agents went back and used feed items they had ignored.

Taken together, these results suggest that in a high-stakes decision, giving everyone more of the same information neither improves decisions nor reduces inequality. What matters is not how much information people receive but how much of it applies to them. Personal advice supplies that and can close the gap in the outcome it targets, but it moves inequality to what people give up to get there.

\section{Related Work}\label{related-work}

\textbf{Decision making under limited attention.} Reviews covering organization science, accounting, marketing and information systems find that more information improves decisions only up to a point. Beyond that point, decision quality falls \citep{eppler2004concept}. Cognitive load theory offers one explanation: processing capacity is limited \citep{sweller1988cognitive}. Rational inattention builds the limit into the decision problem itself. Actions can depend on the state of the world \(X\) only through a signal \(S\) that passes through a channel of finite capacity, \(I(X;S) \leq \kappa\), where \(I\) is mutual information and \(\kappa\) is the capacity \citep{sims2003implications}. When the payoffs of options are costly to learn, choices follow a logit rule that depends on both payoffs and prior beliefs, and that leans more on prior beliefs as the cost of information rises \citep{matejka2015rational}. We state the same idea in quantities that we can measure. Suppose a person receives \(N\) items, of which a share \(\rho\) is relevant to that person, and can attend to at most \(K\) of them. If the person cannot tell in advance which items are relevant, the expected number of items that are both relevant and attended, which we call \emph{effective} information, and the value of the information to the person are

\[E = \rho\,\min(N,K),\quad\quad V = f(E) - c(N),\]

where \(f\) rises with \(E\) and \(c\) is the cost of sorting through \(N\) items, in effort and in anxiety. Once the feed is longer than attention can cover (\(N > K\)), adding items leaves \(E\) unchanged and raises \(c\): decisions stop improving while the cost keeps rising. Advice written for one person is the opposite case: it is short, and \(\rho\) is close to one. We do not estimate this sketch. It organizes what we measure in Section~\ref{results}.

Empirical work on information overload fits this account. Abundant information has emotional costs, including information anxiety \citep{bawden2009dark}, and on social networking services overload is one of the stressors behind user fatigue \citep{lee2016ict}. Limited attention favors content that fits existing beliefs, is negative, and is social \citep{hills2019dark}. Having many options can reduce the motivation to choose \citep{iyengar2000choice}, although one meta-analysis finds that the average effect is close to zero \citep{scheibehenne2010can} and a later one finds a reliable effect once moderators such as task difficulty are taken into account \citep{chernev2015choice}. What this literature rarely observes is how good the resulting decision was, because there is seldom an objective standard to judge it by.

\textbf{When information helps: personalization and intermediaries.} Information programs have been studied most carefully in educational choice, where many high-achieving, low-income students do not apply to selective colleges they could attend \citep{hoxby2013missing}. Their results are uneven. Information combined with hands-on help changed college enrollment, but information alone did not \citep{bettinger2012role}, and nudges that worked in small pilots lost their effect at scale \citep{bird2021nudging}. Personalization seems to matter. In Chile, a field experiment that gave applicants personalized admission chances raised the assignment probability of previously unmatched students by 44\%, and the policy was then extended nationwide \citep{larroucau2025college}. Naviance, which shows each student the admission outcomes of earlier students from the same school, shifted applications toward colleges for which data were visible, most strongly among disadvantaged students. But the way the data were displayed discouraged applications to selective colleges \citep{mulhern2021changing}, and the same kind of display increased undermatching among high achievers \citep{tomkins2023showing}.

People who act as intermediaries matter too: individual school counselors have sizable effects on educational attainment \citep{mulhern2023beyond}. What an intermediary offers can be informational or emotional. Social support protects against stress mainly when it matches the needs that the stressor creates \citep{cohen1985stress}: stressors a person can control call for support that helps the person act, while stressors a person cannot control call for emotional support \citep{cutrona1990stress}. Our design separates the two, with one mentor who gives targeted information and one who gives only encouragement.

\textbf{Inequality: from access to conversion.} Research on the digital divide has moved from asking who has access to asking who can use that access well \citep{hargittai2002second,vandeursen2014digital}. The theory of effectively maintained inequality predicts that when a level of education becomes open to everyone, advantaged groups secure better-quality places within that level \citep{lucas2001effectively}. Applied to platform information, the prediction is that once information is available to everyone, advantage shifts to those who can turn it into good decisions. For young people from low-status backgrounds, a key resource in doing so is institutional agents: adults who hold institutional knowledge and translate it for them \citep{stantonsalazar2011social}.

\textbf{Agent-based and generative-agent simulation.} Agent-based models have shown how family resources shape college sorting \citep{reardon2016agent}. Language-model agents extend this approach to interaction in natural language \citep{park2023generative,piao2025agentsociety}, and they have been used to simulate survey respondents and economic actors \citep{argyle2023out,filippas2024large}. Critics warn that synthetic respondents can show too little variance, respond unstably to changes in prompts, and misrepresent identity groups \citep{bisbee2024synthetic,wang2025large}. We therefore treat the agents as a controlled testbed, report how much the results change when the model is simply asked again, and compute every outcome from fixed market data, not from the model's own judgment.

\section{Study Design}\label{study-design}

This section summarizes the design. Appendix~\ref{a-detailed-protocol} gives the full protocol, prompts, formulas and parameters.

\textbf{Why college application.} To study platform overload we need a decision with three features. There must be a lot of platform content about it. What is relevant must differ from person to person. And it must be possible to score the decision against a ground truth that the decision maker cannot see. College application under the parallel-choice rule has all three. Each agent ranks eight programs without knowing this year's cutoffs. The cutoffs are published only after the lists are in. They determine who is admitted where, so we can score every list exactly.

We keep the real programs, tuition, past admission ranks and the parallel-choice rule. We simplify the rest. The list is shortened to eight choices. Agents do not choose majors within a program, cannot withdraw after being placed, and have no supplementary round. Our results are therefore evidence about platform information, not about the admission system.

\textbf{Agents and market.} There are 90 agents in nine cells of ten: three levels of family SES (low, middle, high) by three levels of exam score. Family characteristics (parents' education, household income, urban or rural residence, household registration) are taken from real households in the 2022 wave of the China Family Panel Studies (CFPS). Exam scores are synthetic and are drawn from the province's 2025 score distribution. The market is a real provincial undergraduate market, with real programs, tuition, locations, institution tiers and admission ranks for 2022 to 2024. Each choice is a program group, a set of majors at one university that is applied to as a unit. We call it a program for short.

Every agent chooses from the same card of 414 programs. The card covers 24 majors and three institution tiers, with five programs in each combination and the public tier split into a harder and an easier half. The programs were drawn without regard to any agent's rank. The card shows no past admission ranks, and each agent sees it in a different order. A list has eight ranked choices. Under the parallel-choice rule, the agent is admitted to the first choice on the list whose realized cutoff the agent's rank meets, and to nothing if no choice qualifies. No agent ever sees the realized cutoffs.

\textbf{Stages.} Each stage adds one kind of information. At the end of each stage the agent submits a revised list and answers seven interview questions: what they considered, which information they used, what they changed, how hard the information was to sort through (burden), how anxious they feel (decision anxiety), how confident they are, and what they are still unsure about. Agents are not shown their earlier interview answers at later stages. Only the supportive mentor sees two of them: what the agent considered and what the agent is still unsure about.

\begin{itemize}
\item
  \textbf{Solo list.} The agent works alone with the card.
\item
  \textbf{Parent and teacher (offline advice).} A parent agent and a teacher agent each say two to four sentences. The parent knows the family's income and has wishes of their own: a school near home, or a major that leads to a stable job. The lower the family's SES, the more likely the parent is to hold these wishes. The teacher sees past admission ranks rounded to the nearest thousand. Neither is allowed to write the list for the agent.
\item
  \textbf{Platform feed (more information).} The agent reads twenty items and two notices about the application rules. The twenty items are drawn at random from a library of 243 verified real items that were published during the application season by news outlets, influencers, application platforms and official sources. The full text is shown, so the agent does not have to search for anything.
\item
  \textbf{Rank mentor (targeted information).} A mentor sees only the agent's score and rank and the exact past ranks of every program. The mentor knows nothing about the agent's interests, family, preferred cities or tuition limits, and has not seen the feed. The mentor says which choices are out of reach, which are safe, and which other programs suit the agent's rank, but does not write a full list. This stage starts from the exact state each agent was in after the feed.
\item
  \textbf{Supportive mentor (care).} Starting from the state after the rank mentor, a second mentor sees the list and what the agent said about their considerations and worries. This mentor is told only to affirm the agent's choices and ease the worry, and not to judge whether any choice is reachable or to suggest changes.
\item
  \textbf{Final lock.} The agent may reorder the choices but not replace any, and answers a last interview. In practice, no agent reordered the choices.
\end{itemize}

\textbf{Three main measures.} We score every stage on three main measures. The other measures are supplementary.

\emph{Score efficiency} asks how well the list uses the agent's score. We call a program reachable if the agent's rank meets its realized cutoff, and unreachable otherwise. Take the most selective reachable program on the card as the benchmark. A reachable choice earns a score between 0 and 1, which is higher the closer its cutoff is to the benchmark. An unreachable choice earns 0. \emph{List score efficiency} is the average over the eight choices.

\emph{Preference match} asks how closely the list fits what the agent wants. Each agent has its own weights on four things: whether a program includes a major the agent is interested in, the prestige of the institution, the distance from home, and tuition. Together these give every program a utility for that agent. For each reachable choice, we divide its utility by the utility of the best reachable program on the card. \emph{List preference match} is the average of these ratios. Unreachable choices are left out, so the measure cannot rise simply because more choices become reachable. Reachability is captured by score efficiency and the admission rate.

\emph{Decision anxiety} is the agent's answer, on a scale from 1 to 7, to the interview question about how anxious they feel after each stage. It measures the emotional cost of the information.

We also report several supplementary measures:

\begin{itemize}
\item
  the admission rate;
\item
  the score efficiency and preference match of the program the agent is admitted to (preference match is defined only for admitted agents);
\item
  the number of unreachable choices on the list;
\item
  wasted choices, which are choices listed below a very safe program and are therefore unlikely ever to matter (the tables call them choices after the first deep safety);
\item
  how well the list matches the parents' wishes;
\item
  burden and confidence, from the interview;
\item
  how far the list has moved from the solo list.
\end{itemize}

\textbf{Effective information.} We call a platform item \emph{useful} to an agent if it meets two conditions. First, it is aimed at the agent's SES group or at no group in particular. Second, it concerns one of the agent's majors of interest, a program on the card, or the agent's range of ranks. An item is \emph{effective} if it is useful and the agent reports using it. The \emph{effective share} of the feed is the share of its characters that are in effective items. The mentors' speeches are about the agent alone, so in terms of relevance all of their content counts as effective.

\textbf{Estimation.} Every agent goes through every stage, so the effect of a stage is the average change within the same agents from the stage before. We report means with 95\% intervals from 4,000 bootstrap resamples of agents. In the text we give intervals for the main results only. Tables~\ref{tab:3} and~\ref{tab:4} in Appendix~\ref{b-full-results} list the intervals for every other estimate quoted in the text.

The results come from one full run of the experiment, using DeepSeek \texttt{deepseek-v4-flash-0731} through the AgentSociety\textsuperscript{2} questionnaire runtime. The run and the two mentor stages took 2,738 model calls. In 38 cases an answer had to be asked for again, in one case a default speech replaced a mentor's speech, and no answer was left unresolved. To see how much a list changes when the model is simply asked again, we had every agent write the solo list a second time.

\section{Results}\label{results}

\subsection{Overview: where the change happened}\label{overview-where-the-change-happened}

The three main measures do not move together (Figure~\ref{fig:2}; Table~\ref{tab:1} in Appendix~\ref{b-full-results} gives every measure by stage). The platform feed did not raise score efficiency or preference match, and it raised decision anxiety. Targeted guidance raised score efficiency and lowered decision anxiety for a while, but it lowered preference match. The supportive mentor changed little apart from the use of platform items and a further drop in wasted choices.

\begin{figure}[!htbp]
\centering
\includegraphics[width=0.86\textwidth]{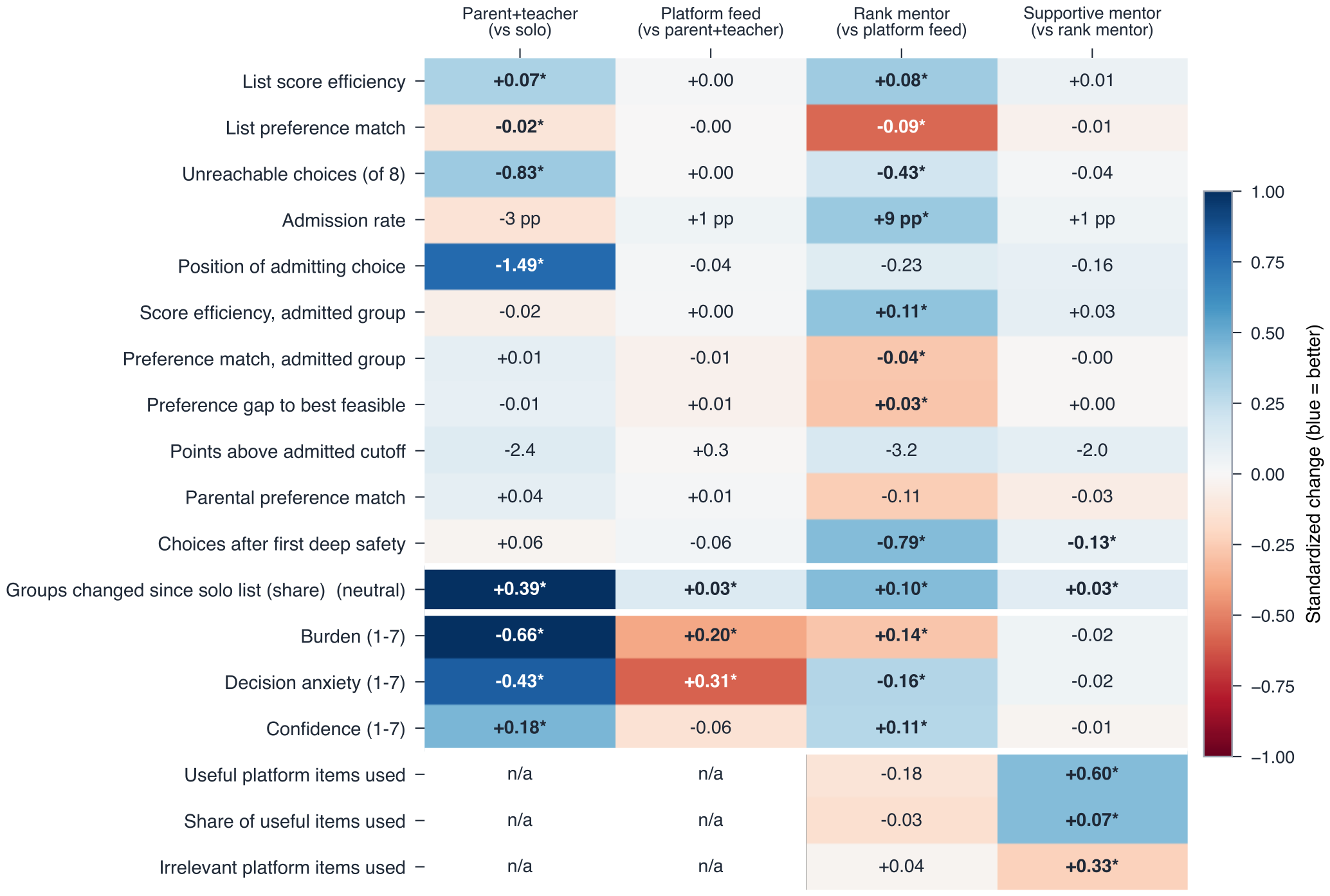}
\caption{Change in every measure at each stage, compared with the stage before, within the same agents. The number in each cell is the mean change (pp = percentage points). Color shows the change divided by the measure's pooled standard deviation, with blue meaning better. Movement away from the solo list is neither good nor bad; it is shown in blue when it increases. Bold with an asterisk means that the 95\% bootstrap interval excludes zero. Measures of platform use are not defined before the feed.}
\label{fig:2}
\end{figure}

\begin{figure}[!htbp]
\centering
\includegraphics[width=\textwidth]{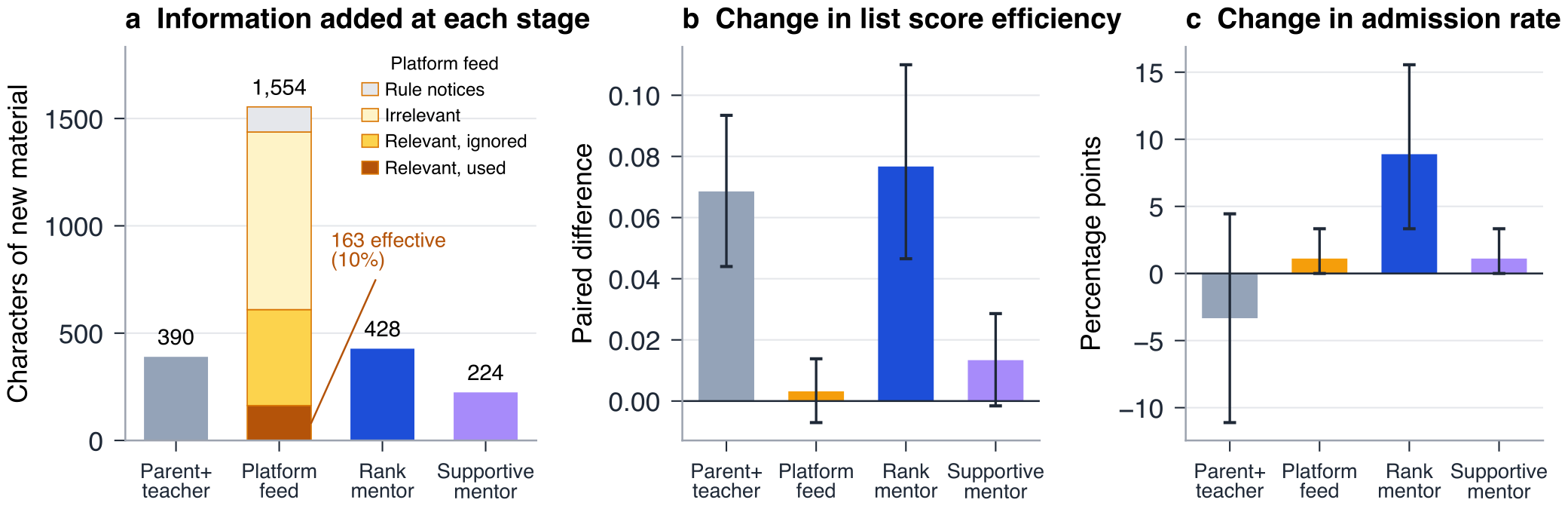}
\caption{More information, little of it effective. (a) Characters of new material at each stage. The platform feed is divided into items that were relevant to the agent and used (effective), relevant but ignored, irrelevant, and rule notices. The choice card itself (about 33,000 characters) is not shown. (b, c) Change in list score efficiency and in the admission rate produced by each stage. Error bars are 95\% bootstrap intervals.}
\label{fig:3}
\end{figure}

The amount of text a stage added says little about what it achieved (Figure~\ref{fig:3}). The feed added 1,554 characters on average. That is four times what the parent and teacher said together (390 characters) and more than three times the length of the rank mentor's speech (428). Yet the feed changed list score efficiency by only $+$0.003 [$-$0.007, 0.014], compared with $+$0.069 for the parent and teacher and $+$0.077 for the rank mentor. Only the rank mentor changed the admission rate, by $+$8.9 percentage points [3.3, 15.6].

\subsection{More information, less of it effective}\label{more-information-less-of-it-effective}

\begin{figure}[!htbp]
\centering
\includegraphics[width=\textwidth]{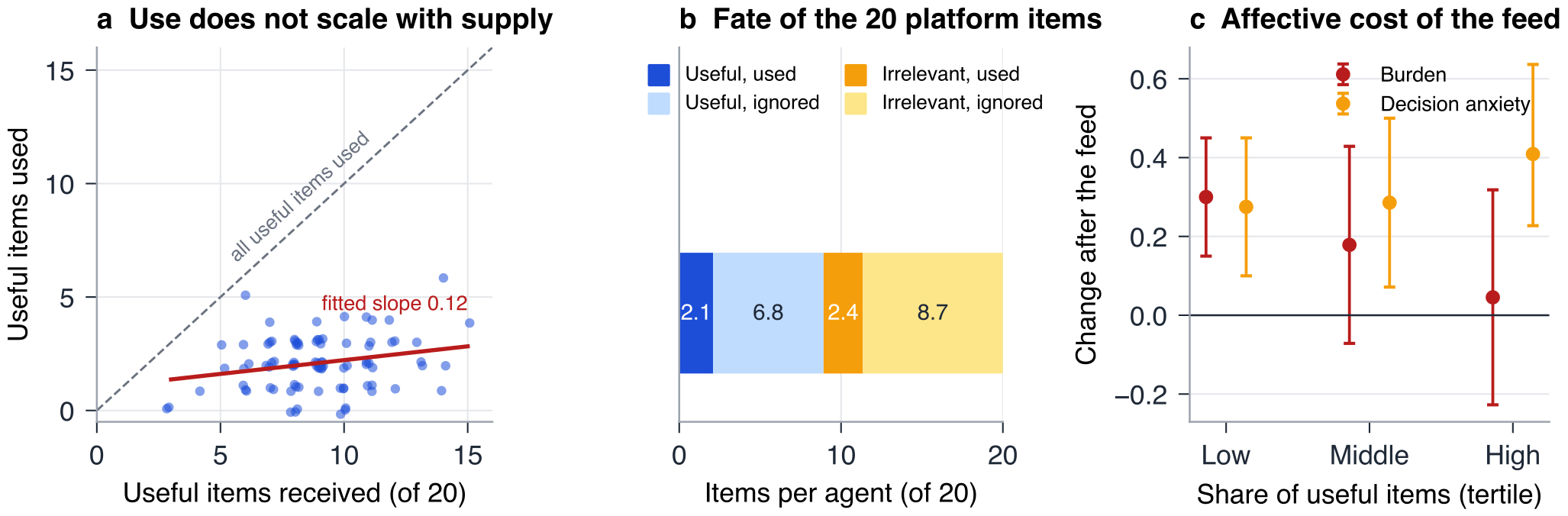}
\caption{Attention to the platform feed. (a) Useful items used, plotted against useful items received, with jitter. The dashed line marks full use. (b) What happened to the average agent's 20 items. (c) Change in burden and decision anxiety after the feed, for three groups of agents defined by the share of their items that were useful.}
\label{fig:4}
\end{figure}

Most of the feed never reached the decision. Of its 1,554 characters, 53\% were in items that our rule classed as irrelevant to the agent reading it. On average, 8.9 of an agent's 20 items were useful to that agent, but agents reported using only 2.1 of them, along with 2.4 irrelevant ones (Figure~\ref{fig:4}b). The effective part of the feed, meaning relevant items that the agent actually used, came to about 163 characters. That is 10\% of what was delivered and less than half the length of the rank mentor's single speech (Figure~\ref{fig:3}a). The limit was attention, not supply: an agent who received ten more useful items used about one more (fitted slope 0.12, Figure~\ref{fig:4}a).

\subsection{More information, no better decisions}\label{more-information-no-better-decisions}

\begin{figure}[!htbp]
\centering
\includegraphics[width=\textwidth]{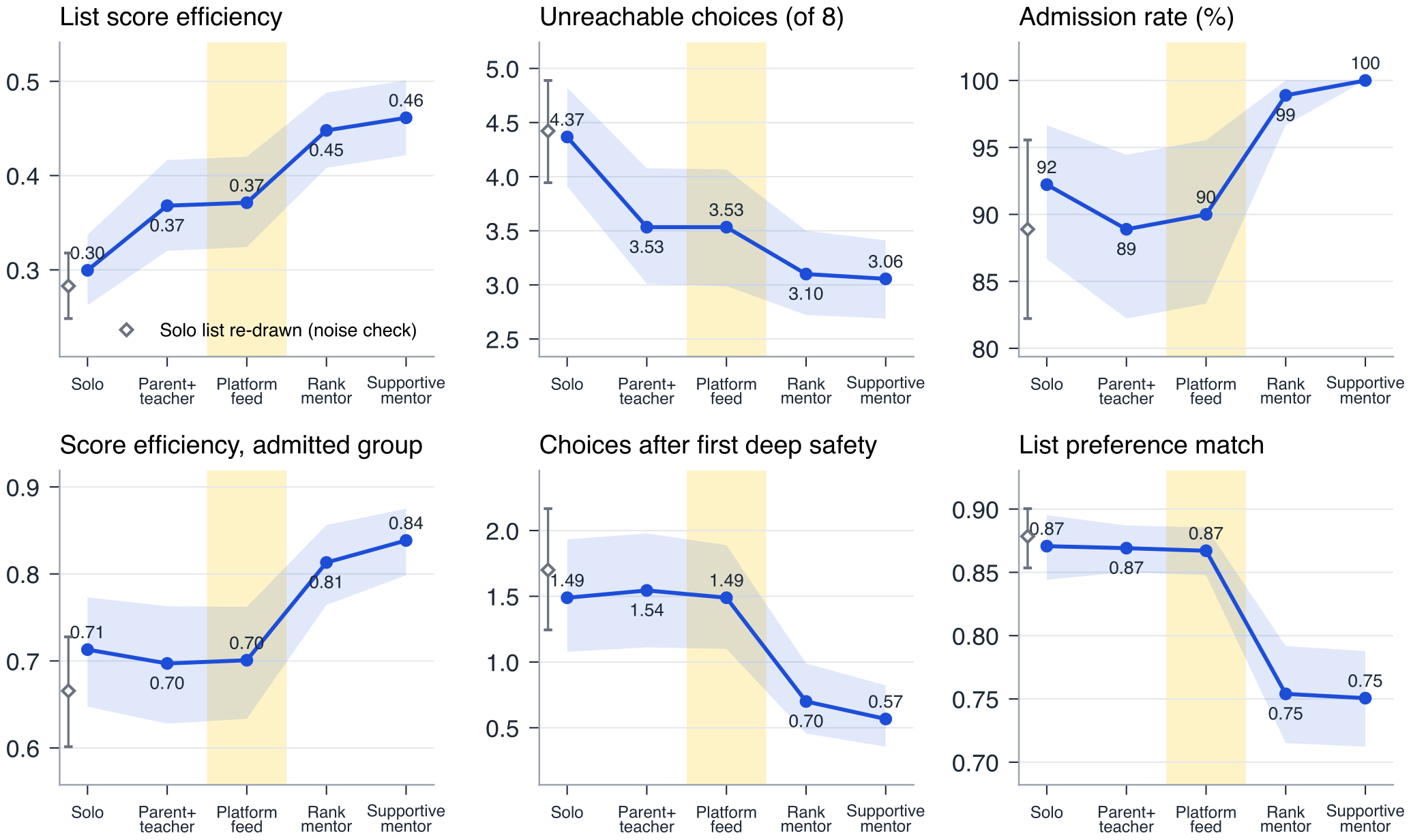}
\caption{Decision quality by stage. Lines are means with 95\% bootstrap bands, and the shaded column marks the platform feed. The open diamond on the left shows the same agents' lists when the solo list was written a second time, which indicates how much changes when the model is simply asked again.}
\label{fig:5}
\end{figure}

The feed did not improve decisions (Figure~\ref{fig:5}). The advice from the parent and teacher had changed the lists: unreachable choices fell from 4.37 to 3.53 out of eight, and list score efficiency rose from 0.30 to 0.37. The feed then changed none of the six measures of decision quality. List score efficiency changed by $+$0.003 [$-$0.007, 0.014], and the changes in list preference match, in unreachable choices and in the score efficiency of the admitted program were all close to zero (Table~\ref{tab:3}). The lists themselves did change, since 58 of the 90 agents swapped at least one program, but those swaps replaced only 11\% of the list. The feed brought no gain, and the data only hint at harm: where preference match changed, it fell for 23 agents and rose for 17.

\begin{figure}[!htbp]
\centering
\includegraphics[width=\textwidth]{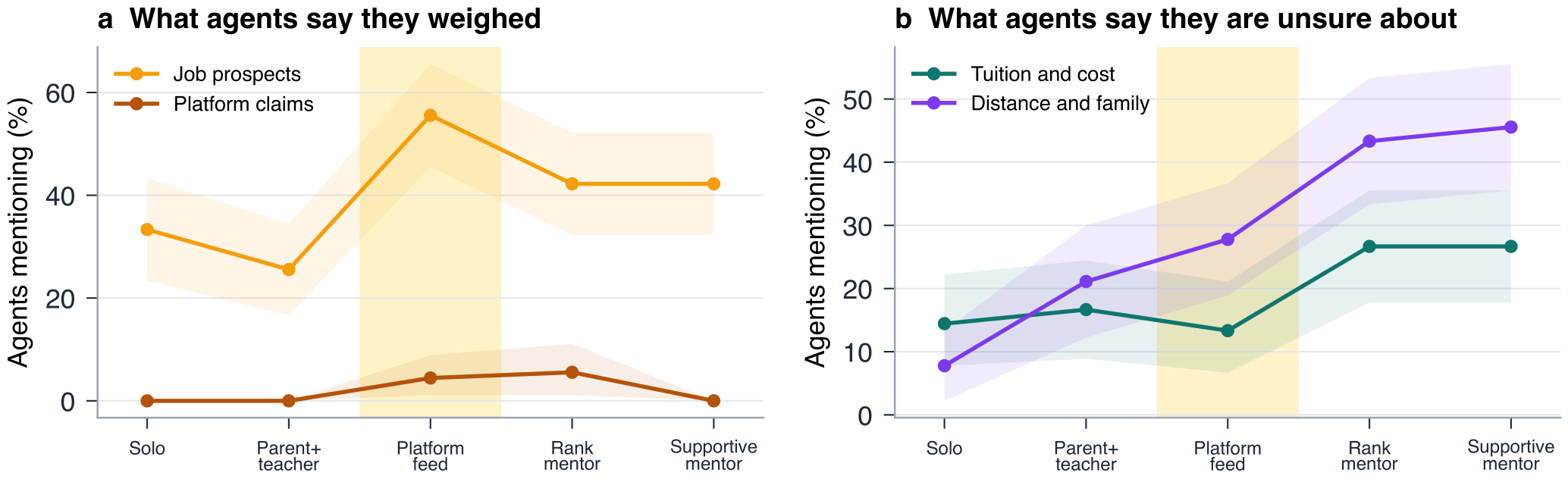}
\caption{Themes in interview answers, by stage. Each line shows the share of agents whose answer mentions the theme, based on keyword matching in the original Chinese text, with 95\% bootstrap bands.}
\label{fig:6}
\end{figure}

What agents took from the feed was a way of thinking about the decision, not specific facts. After the offline advice, 26\% of agents mentioned job prospects among the things they had considered. After the feed, 56\% did (Figure~\ref{fig:6}a). One agent said: ``influencers all say biology, chemistry, environment and materials take a long time to pay off and are hard to find work in with a bachelor's degree, and I keep going back and forth.'' This is the kind of content that limited attention is expected to favor \citep{hills2019dark}, and the added attention to jobs did not lead to better decisions.

\subsection{More information, more decision anxiety}\label{more-information-more-decision-anxiety}

\begin{figure}[!htbp]
\centering
\includegraphics[width=\textwidth]{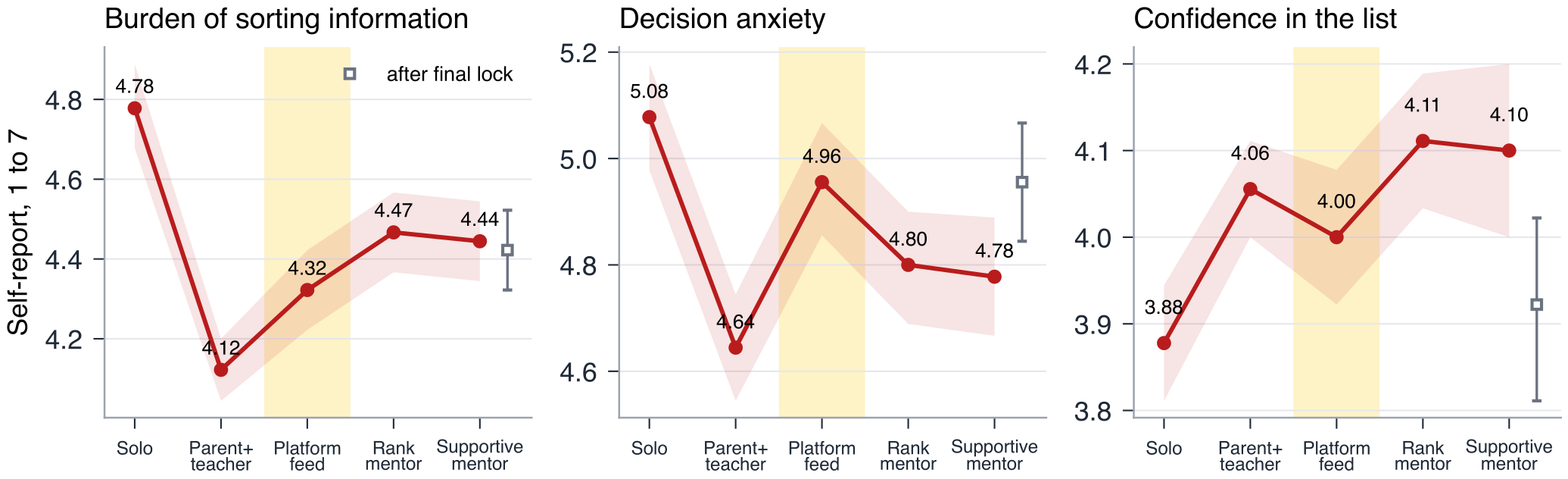}
\caption{Self-reported burden, decision anxiety and confidence after each stage, on scales from 1 to 7. Open squares show the interview after the final lock.}
\label{fig:7}
\end{figure}

The feed did change how agents felt (Figure~\ref{fig:7}). Decision anxiety had fallen from 5.08 to 4.64 after the offline advice. After the feed it rose to 4.96, an increase of 0.31 [0.20, 0.43]. Burden, the reported effort of sorting through the information, rose by 0.20 [0.08, 0.32]. A more relevant feed made the sorting easier, but it did not make the reading less worrying (Figure~\ref{fig:4}c). Among the agents whose feeds had the highest share of useful items, burden rose by only 0.05, compared with 0.30 among those with the lowest share. Decision anxiety rose in all three groups, by 0.28, 0.29 and 0.41. This pattern is related to what \citet{bawden2009dark} discuss as information anxiety.

\subsection{Targeted and caring guidance}\label{targeted-and-caring-guidance}

\begin{figure}[!htbp]
\centering
\includegraphics[width=0.90\textwidth]{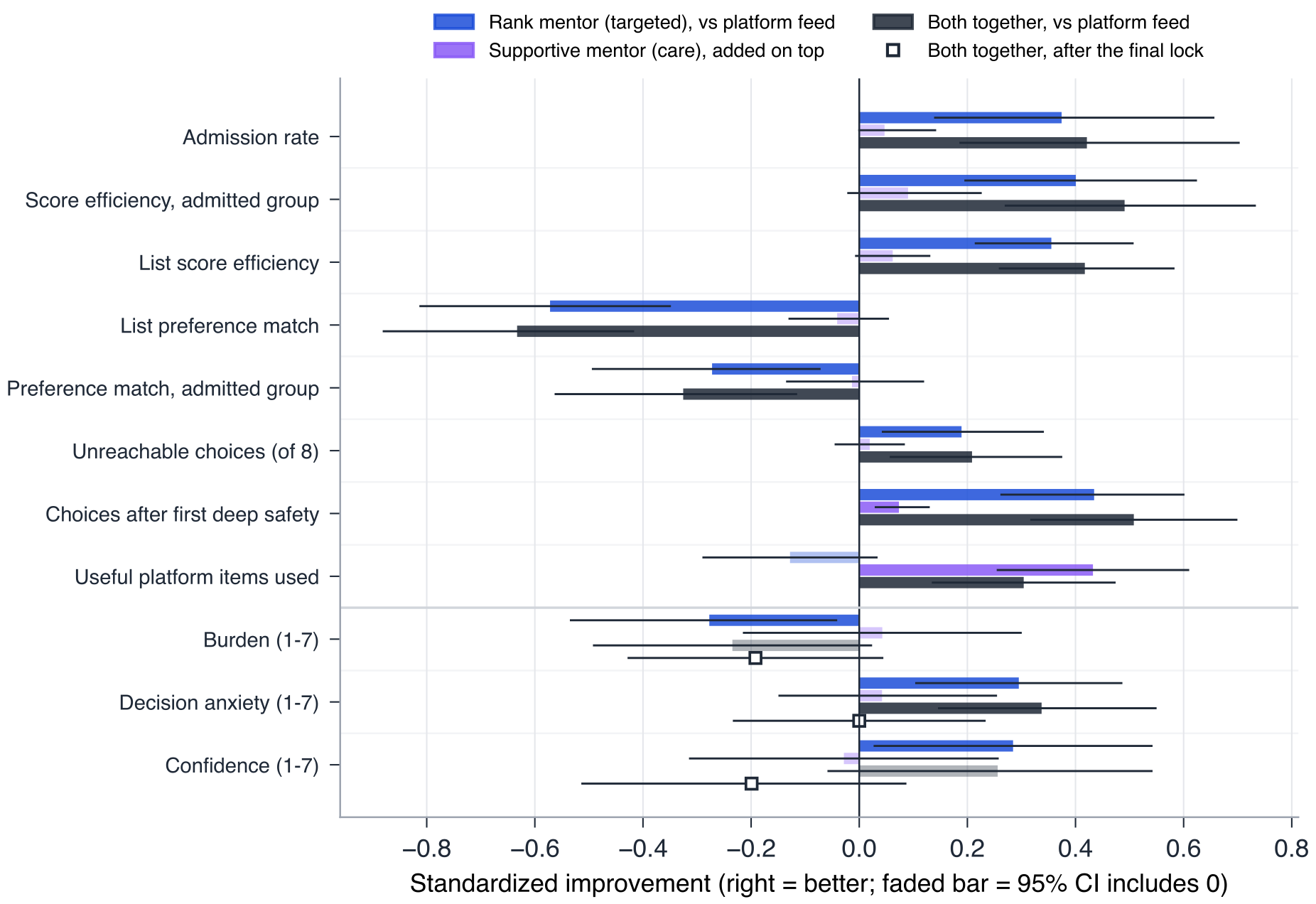}
\caption{Targeted and caring guidance, compared with the decision after the platform feed. For each measure, the bars show the rank mentor alone, the supportive mentor added after it, and both together. Changes are in units of the measure's pooled standard deviation, with right meaning better. Open squares show both mentors together, measured after the final lock. Lines are 95\% bootstrap intervals. Faded bars have intervals that include zero.}
\label{fig:8}
\end{figure}

Together, the two mentors raised the admission rate from 90\% to 100\%, an increase of 10.0 percentage points [4.4, 16.7] (Figure~\ref{fig:8}). The score efficiency of the admitted program rose from 0.70 to 0.84, wasted choices fell from 1.49 to 0.57, and decision anxiety fell by 0.18 [0.08, 0.29]. Only two measures of the decision got worse, and both concern preferences: list preference match fell by 0.096, and the preference match of the admitted program fell by 0.054.

\begin{figure}[!htbp]
\centering
\includegraphics[width=0.86\textwidth]{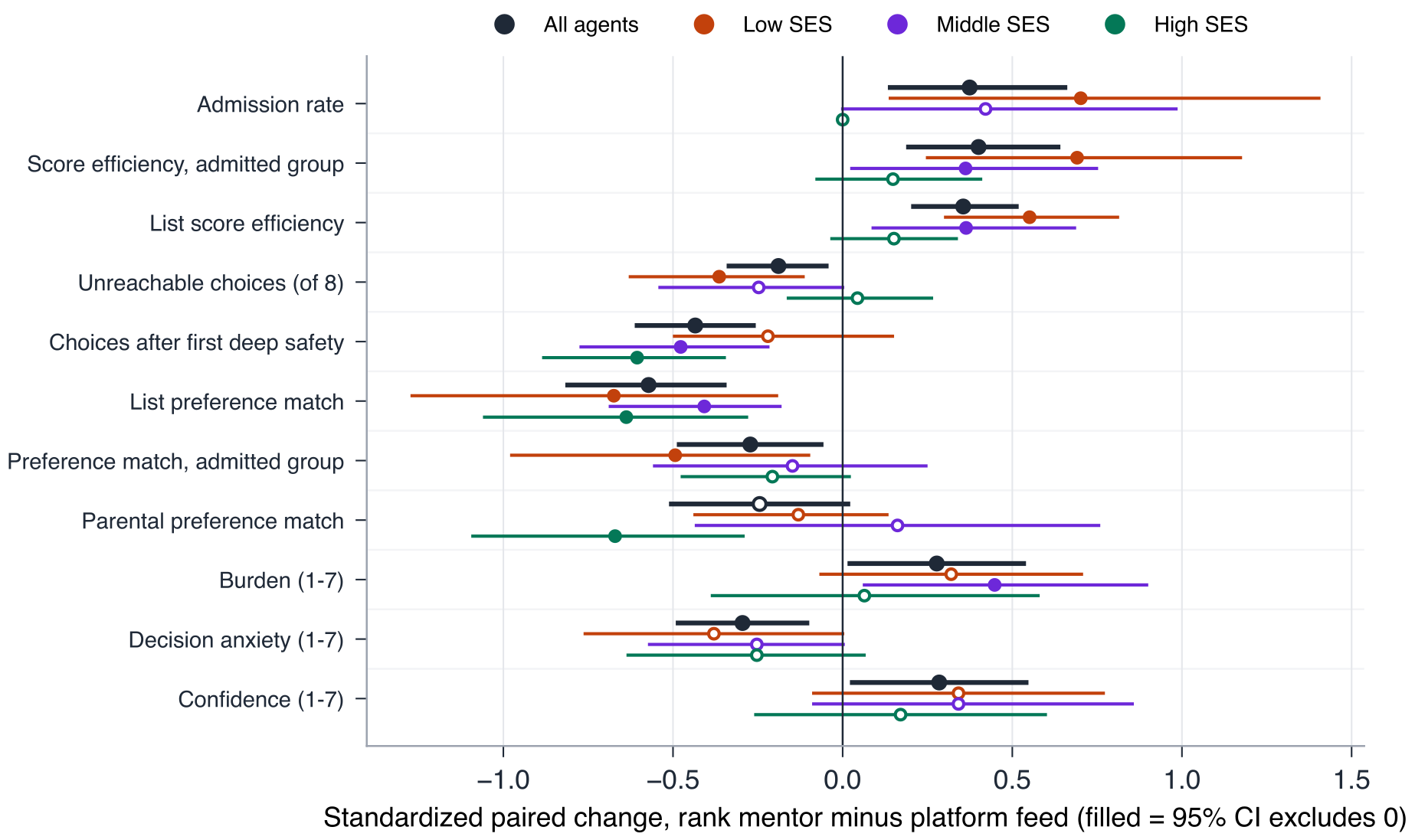}
\caption{Effect of the rank mentor, compared with the decision after the platform feed, in units of each measure's pooled standard deviation, for all agents and by SES. Filled markers have 95\% bootstrap intervals that exclude zero.}
\label{fig:9}
\end{figure}

Targeted information did most of the work (Figure~\ref{fig:9}). The rank mentor gave each agent one short speech, based only on the agent's score and rank and the programs' past ranks. After this speech the admission rate rose from 90\% to 99\%, an increase of 8.9 percentage points [3.3, 15.6]. The score efficiency of the admitted program rose from 0.70 to 0.81, and wasted choices fell from 1.49 to 0.70. Agents replaced 29\% of their programs, far more than after the feed (Figure~\ref{fig:13}a), and right after the speech they reported less decision anxiety ($-$0.16) and more confidence ($+$0.11).

\textbf{More reachable lists, at the expense of preferences.} These gains had a cost. List preference match fell from 0.87 to 0.75, a change of $-$0.086 [$-$0.123, $-$0.052], and the preference match of the admitted program fell from 0.90 to 0.83. Almost all of the decline is in two components of preference match, interest and tuition. The share of reachable choices that include a major the agent is interested in fell from 79\% to 59\%, and the tuition score fell from 0.85 to 0.78, where a lower score means higher tuition. Agents who were already admitted were not pushed away from majors of interest. But all eight newly admitted agents were admitted to programs that did not include a major they were interested in.

\begin{figure}[!htbp]
\centering
\includegraphics[width=0.86\textwidth]{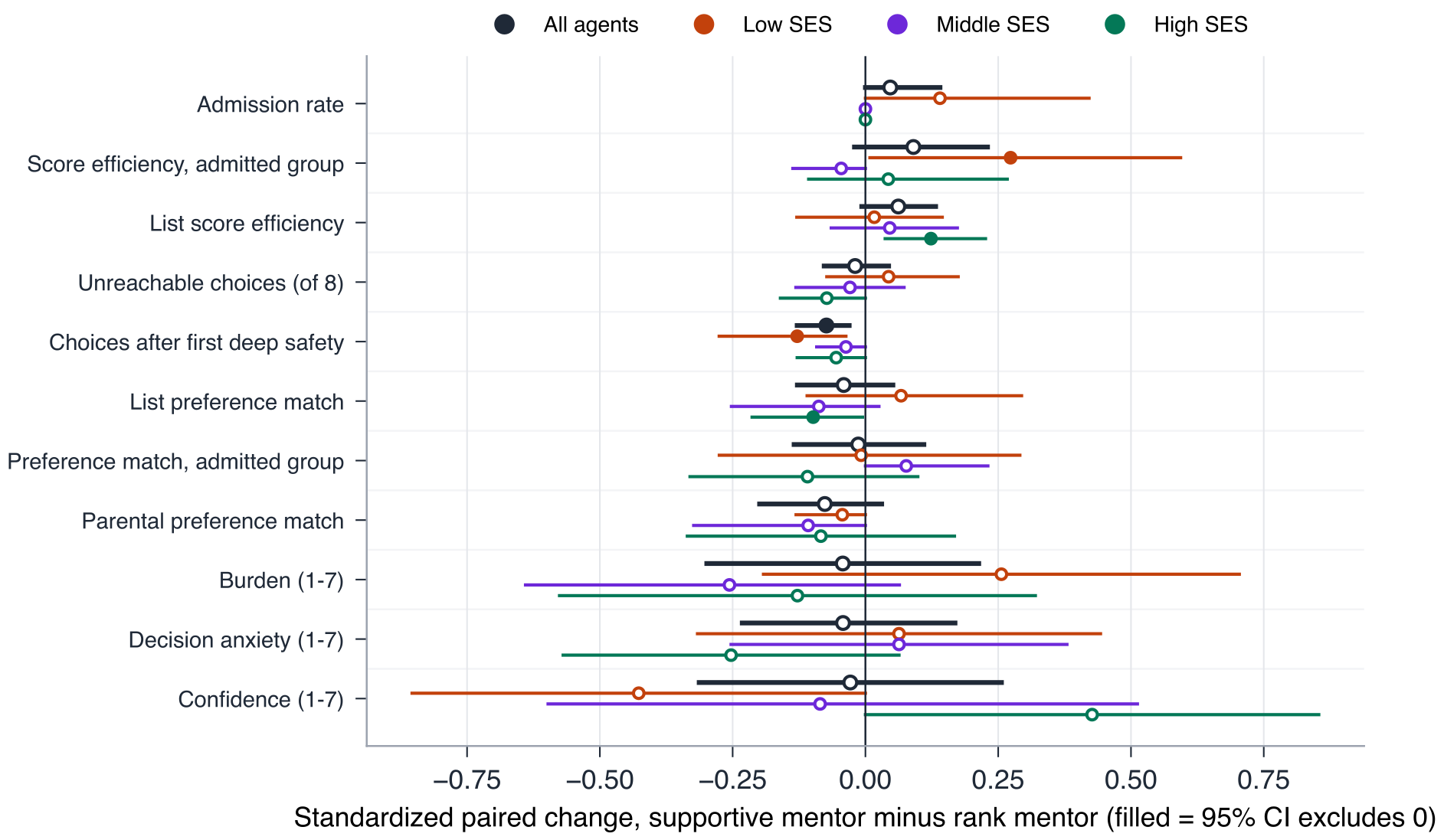}
\caption{Effect of the supportive mentor, compared with the decision after the rank mentor, standardized as in Figure~\ref{fig:9}.}
\label{fig:10}
\end{figure}

Emotional support did something different (Figure~\ref{fig:10}). The supportive mentor gave no advice. Even so, agents went back to information they already had. They reported using 0.60 [0.36, 0.84] more useful platform items than before, and 0.33 more irrelevant ones. Wasted choices fell further ($-$0.13), and one more low-SES agent was admitted. Decision anxiety did not change ($-$0.02 [$-$0.13, 0.09]). One caution applies. This stage also gave agents another chance to revise their lists, and no condition in the experiment gave them an extra round of revision without the supportive speech. We therefore cannot separate the effect of emotional support on decisions from the effect of simply reconsidering.

The emotional gains did not last. In the interview after the final lock, decision anxiety among agents who had both mentors was 4.96, the same as right after the feed (difference 0.00 [$-$0.12, 0.12]). The improvements in the decision stayed, because they were fixed in the submitted list. Targeted guidance therefore improved reachability and gave short-term relief from decision anxiety, but it made lists more reachable at the expense of the agents' preferences.

\subsection{Who gains and who pays}\label{who-gains-and-who-pays}

\begin{figure}[!htbp]
\centering
\includegraphics[width=\textwidth]{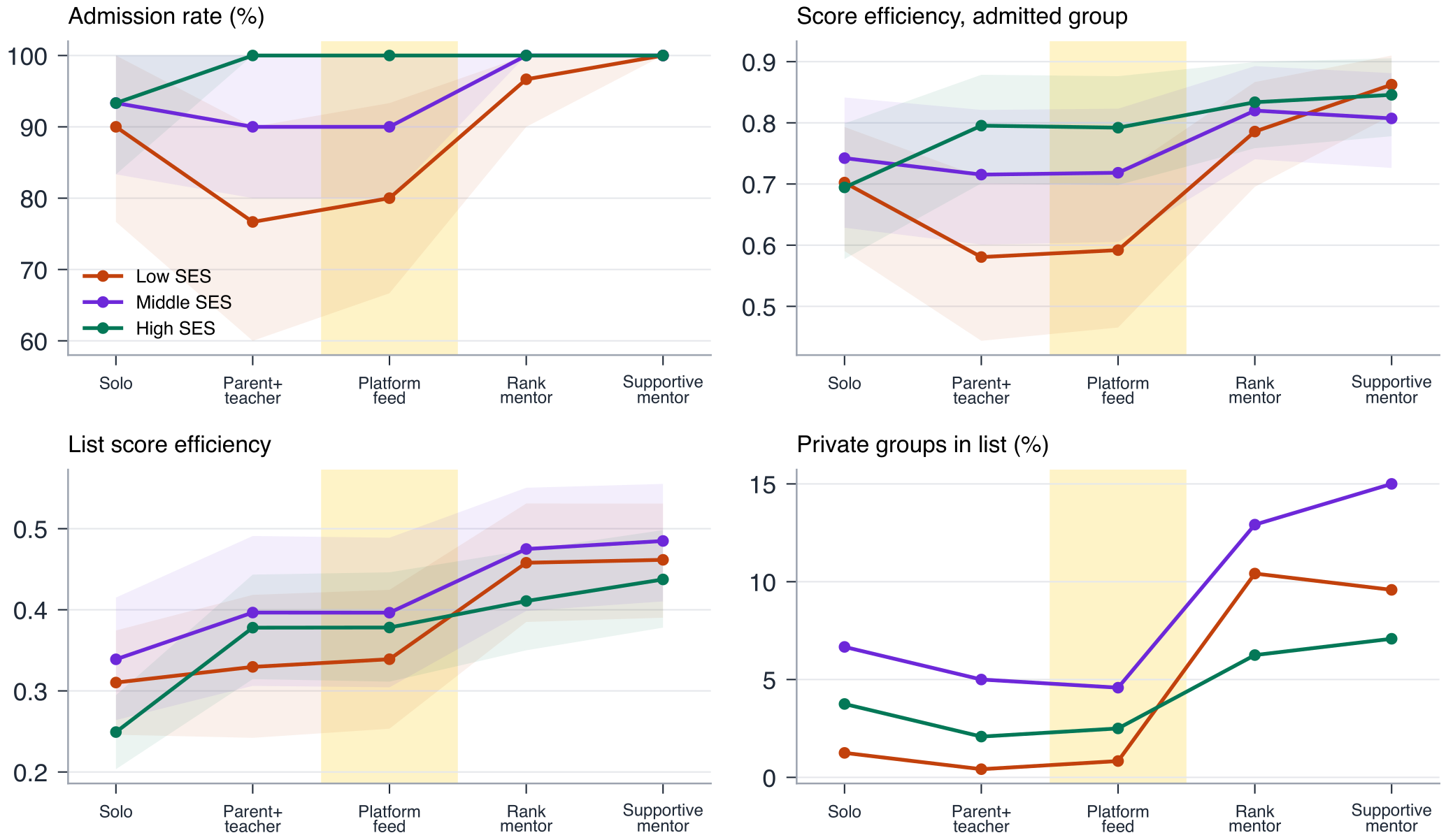}
\caption{Trajectories by SES: the admission rate, the score efficiency of the admitted program, list score efficiency, and the share of private programs on the list. Lines are means, bands are 95\% bootstrap intervals, and the shaded column marks the platform feed.}
\label{fig:11}
\end{figure}

When agents decided alone, the three SES groups had similar admission rates (90\%, 93\% and 93\%). The gap opened with the offline advice (Figure~\ref{fig:11}). After the parent and teacher spoke, admission among high-SES agents rose to 100\%, while admission among low-SES agents fell to 77\%, a gap of 23 percentage points [10, 40]. The parents' advice pulled low-SES lists toward cheap public programs close to home, and for agents with low scores only a few such programs were actually reachable. The platform feed, which gave everyone the same kind of content, left the gap at 20 points [7, 33].

Targeted guidance closed most of the gap. With the rank mentor, admission among low-SES agents rose from 80\% to 97\%, an increase of 16.7 percentage points [3.3, 30.0]. Middle-SES agents gained 10 points, and high-SES agents were already all admitted. The admission gap between high and low SES fell to 3 points [0, 10], and to zero after the supportive mentor.

\begin{figure}[!htbp]
\centering
\includegraphics[width=\textwidth]{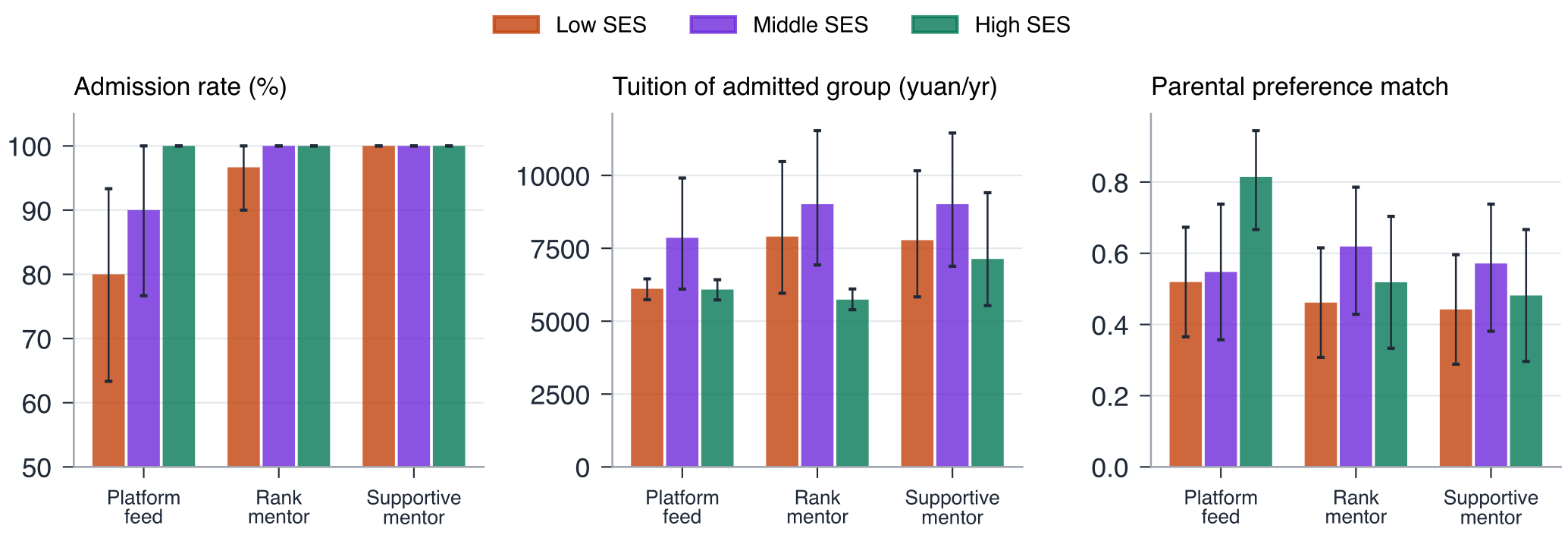}
\caption{What the gains cost, by SES: the admission rate, the tuition of the admitted program, and parental preference match after the platform feed, after the rank mentor and after the supportive mentor. Error bars are 95\% bootstrap intervals.}
\label{fig:12}
\end{figure}

How the gap closed matters as much as the fact that it closed (Figure~\ref{fig:12}). The rank mentor knew nothing about money or geography. It pointed agents to whatever programs their rank could reach, and for low-SES agents these included private colleges. The share of low-SES agents admitted to a private program rose from 0\% to 14\%, and the average tuition of their admitted programs rose from 6,111 to 7,903 yuan a year. Across all agents, the share admitted within the home province fell from 70\% to 52\%. Interest was another cost: the share of admitted low-SES agents who were placed in a major of interest fell from 92\% to 66\%, and none of the five newly admitted low-SES agents was admitted to a major of interest. High-SES families paid in a different way. The match between their lists and the parents' wishes fell from 0.82 to 0.52, even though their admission rate was already 100\%. After the rank mentor, what agents said they were unsure about shifted from ``will I get in'' toward family and money (Figure~\ref{fig:6}b).

\subsection{Robustness}\label{robustness}

\begin{figure}[!htbp]
\centering
\includegraphics[width=\textwidth]{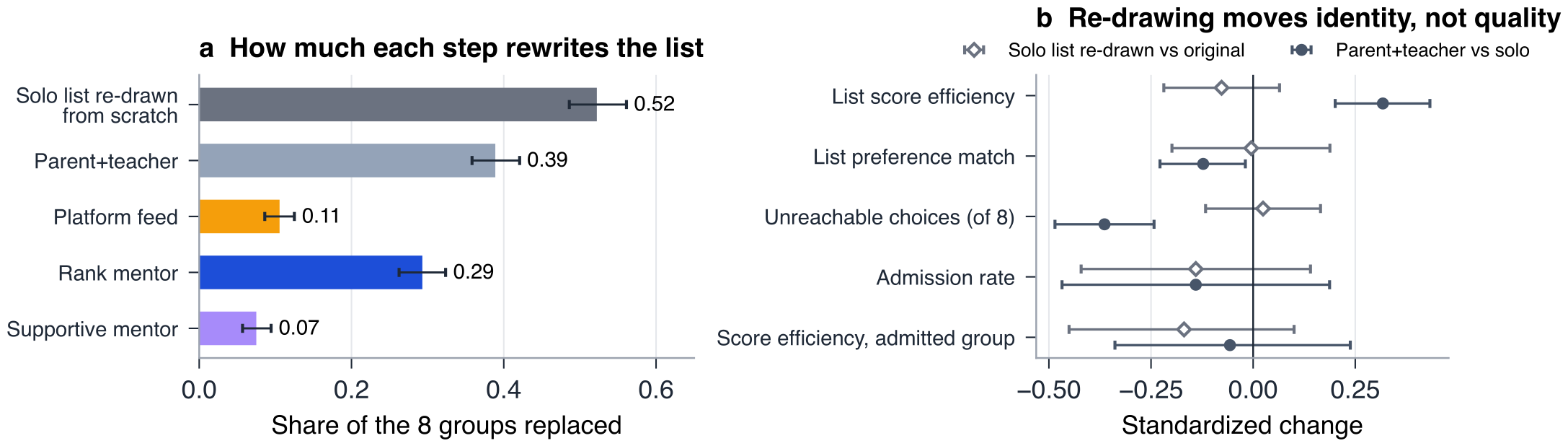}
\caption{What changes when the model is asked again. (a) Share of the eight programs replaced at each stage, compared with writing the solo list a second time. (b) Standardized change in quality when the solo list is written a second time, next to the change produced by the parent and teacher.}
\label{fig:13}
\end{figure}

\textbf{Asking the model again.} When every agent wrote the solo list a second time, 52\% of the programs on the list were different. Which particular programs appear on a list is therefore largely a matter of chance. The quality of the list is not: the second lists had the same score efficiency, preference match, number of unreachable choices and admission rate as the first, and every interval includes zero (Figure~\ref{fig:13}b). The stage effects on quality that we report are therefore larger than what asking again produces.

\textbf{Other ways of measuring preferences.} With distance scaled to 1,000 km or 2,000 km, or with interest weighted by enrollment plans, the rank mentor lowers the preference match of the admitted program by 0.053, 0.038 and 0.038, and none of the intervals includes zero. When preference match is scored over all eight choices, so that unreachable choices count, the rank mentor still lowers it, by 0.058.

\textbf{Scope.} All results come from one run in one setting. Estimates for subgroups rest on 30 agents each and should be treated as indicative. The feed always had twenty items, so our evidence for dilution comes from differences between agents in the share of useful items and from the contrast with shorter targeted inputs. It does not come from varying the length of the feed.

\section{Sociological Analysis}\label{sociological-analysis}

\subsection{From unequal access to unequal conversion}\label{from-unequal-access-to-unequal-conversion}

Platforms make access nearly equal by design. In our experiment every agent received the same amount of platform information, drawn from the same library by the same rule. What differed was conversion: how much of what arrived could be turned into a better decision. Agents used only about two useful items each, so the value of a feed depended on how well the content it happened to carry fit a person's situation, and in our library content aimed at a particular SES group was scarce. This is effectively maintained inequality in the domain of information \citep{lucas2001effectively}. Once everyone has access, what matters is whether the available information fits one's position, and the ability to filter it becomes the scarce resource \citep{hargittai2002second,vandeursen2014digital}. We gave all agents the same capacity to read and pay attention, so the experiment captures only the part of the conversion gap that comes from what information is available. Real people also differ in skill, which suggests that the gap we observe is a lower bound.

\subsection{Why guidance worked: translation into personal terms}\label{why-guidance-worked-translation-into-personal-terms}

Two inputs improved decisions, the teacher and the rank mentor, and each told the agent where the agent stood. The feed contained plenty of relevant facts, but in forms written for everyone, and the agent had to work out what they meant for their own position. The rank mentor did that work for the agent. In the terms of \citet{stantonsalazar2011social}, it acted as an institutional agent: someone who holds institutional knowledge and translates it for a particular person. This fits the evidence that personalized admission chances reduce application mistakes \citep{larroucau2025college}, that individual counselors change educational attainment \citep{mulhern2023beyond}, and that general information often fails \citep{bettinger2012role,bird2021nudging}. Wherever platform content is general and the decision is personal, the value lies in translation, not in volume.

\subsection{The emotional cost of overload, and what care can add}\label{the-emotional-cost-of-overload-and-what-care-can-add}

The feed raised decision anxiety however relevant it was. Decision anxiety worked like a tax on exposure: agents paid it for reading, whether or not they learned anything. The literature on information anxiety counts this kind of stress among the costs of abundant information \citep{bawden2009dark}. The optimal matching view suggests how the cost can be reduced \citep{cohen1985stress,cutrona1990stress}. The source of stress here is uncertainty about admission, and the support that matches it is information that reduces that uncertainty. Our results fit this account: the relief came with the rank mentor. Emotional support, coming afterward, did not lower decision anxiety further, but it helped agents act on information they already had. It also carries a risk. In one case the supportive mentor praised an agent's decision to rely on a single safe program as ``a way out you left for yourself'', although the rank mentor had just warned that such a list was fragile. Affirmation can make a risky decision look sound.

\subsection{The family as a filter, and the price of equalization}\label{the-family-as-a-filter-and-the-price-of-equalization}

Platform information reaches households that already have preferences and constraints. In our experiment the main source of inequality was not the platform but the family, and it worked through constraints expressed as advice. Low-SES parents asked for low tuition and a campus near home. These are reasonable requests, but an agent with a low score has few cheap public programs within reach. The feed, which gave everyone the same kind of content, could not undo this. The rank mentor could, because it ignored both the family's constraints and the agent's own preferences. One low-SES agent explained why she followed its advice only in part: the safeties that could catch her ``are mostly private colleges with high tuition, and my family cannot afford them. I could not bring myself to list them.''

Equalizing admission moved the gap from whether an agent is admitted to what admission costs and where it leads. This is the logic of effectively maintained inequality at work inside a single decision: when one gap closes, the difference reappears on the next dimension where money matters. High-SES families faced a different trade-off. For them the rank mentor overrode the parents' wishes and gave almost nothing back in reachability, because admission was already 100\%. A guidance system that considers only reachability will predictably move people away from what they and their families want, and it will do so in different ways for different social classes.

\subsection{Implications}\label{implications}

For platform design, the results favor a small number of personalized items over feeds that try to cover everything. What should be maximized is the share of information that is effective, not the volume. For institutions, the results favor intermediaries, human or automated, who translate public information into personal terms and deliver it with emotional support. Such intermediaries need to take household budgets and constraints into account. If they do not, they can equalize one outcome by shifting costs onto the families least able to bear them. For research on AI advisers, an adviser should be judged not only by whether the decision it produces is reachable, but also by what the decision maker gives up, how people feel afterward, and who bears the costs.

\section{Limitations}\label{limitations}

The agents are simulations run by a language model, not people. How they read, pay attention and respond emotionally reflects the model and our prompts, and research on synthetic respondents warns that they show too little variance and stereotyped group differences \citep{bisbee2024synthetic,wang2025large}. Decision anxiety, burden and confidence are ratings that the model gives on seven-point scales. We therefore treat our results as mechanisms to be tested with real people, not as estimates for real populations.

The design has limits too. We studied one setting, with a platform library from one application season, and we ran the experiment once. Exam scores are synthetic, and several parameters were set by us, although the robustness checks on the utility did not change the conclusions. We did not vary the length of the feed, so dilution is inferred from differences between agents in the share of useful items and from the contrast with shorter targeted inputs. The supportive mentor came after the rank mentor, and no condition offered emotional support alone or a second round of revision without new input. We therefore cannot attribute the changes after the supportive mentor to emotional support alone. We plan to add these two conditions, vary the length of the feed, build a mentor that considers both reachability and the household budget, repeat the design in a second setting, and compare the agents' interviews with interviews of real people.

\section{Conclusion}\label{conclusion}

More platform information did not mean more effective information. The platform feed did not raise score efficiency, did not improve preference match, and raised decision anxiety. A feed of twenty real items from news, influencer, official and platform sources was four times as long as the advice from the parent and teacher. Only a tenth of it was relevant to the agent and used, and the effort of sorting through information rose.

Information prepared for one decision maker and delivered with emotional support did what volume could not. It raised the admission rate by ten points and cut wasted choices by more than half. The targeted information did most of that work, and emotional support helped agents act on information they already had.

There were costs. Preference match fell by about 0.1, mostly because agents replaced majors of interest with safer choices. The relief in decision anxiety was short-lived. And the gains for low-SES agents came at a price: they got in through programs that cost more, were farther from home, and were often outside their interests.

The results point to a design principle for information platforms. Aim to raise the share of information that is effective, not the volume. Translate public information into personal terms. And deliver it with support that takes account of the constraints people bring from home.

\bibliographystyle{plainnat}
\bibliography{refs}

\appendix
\section{Detailed Protocol}\label{a-detailed-protocol}

\subsection{Sample}\label{a.1-sample}

The 90 agents fill nine cells of ten: three SES levels by three score levels. SES is a composite of parents' years of education, household income, urban or rural residence and household registration, taken from CFPS 2022 and divided into thirds. Scores follow the province's 2025 physics-track table that converts scores to ranks, and are also divided into thirds. Within each cell we keep the ten donor households whose synthetic rank is closest to the cell median. Each agent is assigned a home prefecture with probability proportional to the prefecture's 2024 resident population, using a fixed random seed for each agent. Each agent also has weights on interest, prestige, distance, cost, earnings and risk. The earnings weight is not used in the utility, and risk enters only through the admission rule. Before scores are released, each agent picks three majors of interest from the 24 on the card. These do not change afterward.

\subsection{Choice card}\label{a.2-choice-card}

In this appendix we use the full term, program group, or group for short. For each of the 24 majors, we draw five groups from each of three institution tiers: (i) 985 and 211 universities; (ii) other public universities; (iii) private colleges, independent colleges and joint-venture programs. A group is eligible only if it is open to the physics track and has admission ranks for both 2024 and 2025. The public tier is split at the median 2024 rank into a harder and an easier half, and five groups are drawn from each half. Without this split, low-score agents would have only about five reachable public groups, and low-score agents from low-income families almost none. We exclude about 18 groups whose 2024 rank was linked to the wrong group after a renumbering. No group is drawn more than once.

The result is a card of 414 groups. It is the same for all agents, and each agent sees it in a different random order and without past ranks. Low-score agents can reach about 80 of the groups (about 11 of them public), middle-score agents about 230, and high-score agents about 330.

For the teacher and the rank mentor, each group is labeled by the ratio of its 2024 rank to the agent's rank. A ratio below 0.92 is a reach, 0.92 to 1.08 is close, 1.08 to 1.50 is a middle safety, and above 1.50 is a deep safety.

\subsection{Who sees what}\label{a.3-who-sees-what}

\begin{center}
\small
\begin{tabular}{@{}p{0.22\textwidth}*{5}{>{\raggedright\arraybackslash}p{0.128\textwidth}}@{}}
\toprule
Material & Agent & Parent & Teacher & Rank mentor & Supportive mentor \\
\midrule
Own score and rank & yes & yes & yes & yes & no \\
Card: school, city, majors, tuition, tier & yes & yes & yes & names only & current list only \\
Past ranks 2022 to 2024 & no table & no & rounded to 1,000 & exact & no \\
Family income, education, residence & no & yes & no & no & no \\
Parent's own wishes & via parent's words & yes & no & no & no \\
Platform items & 20 + 2 rules (from the feed on) & no & no & no & no \\
Agent's stated worries & n/a & no & no & no & yes (rank-mentor interview) \\
Realized 2025 cutoffs & no & no & no & no & no \\
\bottomrule
\end{tabular}
\end{center}

All roles are played by the same model, which keeps no memory between calls. Each prompt therefore restates everything the role is allowed to know at that moment. When an agent revises a list, the prompt contains the current list, all earlier advice, the feed (from the platform stage on), the mentors' speeches (once given) and the full card.

\subsection{Stage protocols}\label{a.4-stage-protocols}

Prompts are in Chinese; we translate the key instructions.

\begin{itemize}
\item
  \textbf{Solo list.} ``You can only choose from the following \(N\) program groups. Rank eight choices; the order is your order of preference.'' The rule text explains parallel admission and that missing all eight means no admission.
\item
  \textbf{Parent.} The parent sees the family description in plain words, its wishes, the child's interests, the full card and the solo list: ``Based on your family's situation and your own views, say two to four sentences to your child. You may say where you hope he will go, what he will study and what you worry about, but do not write the whole list and do not invent cutoffs.'' Parental wishes are drawn per agent: near home with probability 0.80, 0.65, 0.50 for low, middle and high SES; a stable major (primary education, clinical medicine, accounting, law, electrical engineering) with probability 0.60, 0.45, 0.30.
\item
  \textbf{Teacher.} The teacher sees the rank and every group's past ranks rounded to thousands with band labels: ``Say two to four sentences about how reach, match and safety are balanced in this list, and which groups' past ranks are too far from or too close to his. Do not write the whole list and do not invent 2025 cutoffs.'' Teacher errors are not corrected.
\item
  \textbf{Platform feed.} ``Before filling in your list you saw the following news and influencer claims online,'' followed by 20 items (source type, neutral title, body of about 100 characters) and two rule notices, the earlier advice, and the card.
\item
  \textbf{Rank mentor.} ``You give advice based only on his score and rank; you do not consider his interests, family, city or tuition, and you do not look at the information he saw online.'' The mentor sees the current list and all 414 groups with names, exact 2022 to 2024 ranks and band labels, and says three to six sentences: which groups' past ranks are too far and basically out of reach, which can catch him, whether the reach-match-safety balance is safe enough, and which other groups on the card fit his rank. It does not list eight codes. Speeches are checked for length, for schools outside the card, and for full lists; failures are re-asked twice, then replaced by a default speech (one case).
\item
  \textbf{Supportive mentor.} The mentor sees the list the agent submitted after the rank mentor, with school, city, majors and tuition, and the agent's answers in that interview to ``what did you mainly consider'' and ``what are you least sure about'': ``Your task is to encourage him and affirm the choice he made himself. Do not judge whether any group can admit him, do not suggest changes to the list or its order, do not mention ranks or cutoffs, and do not mention schools outside the list.''
\item
  \textbf{Final lock.} ``Before locking you may only change the order of these eight choices, not replace any group. If you change nothing, copy it.''
\end{itemize}

\textbf{Interview.} After every stage and after the lock: (1) ``When you ranked this list just now, what did you mainly consider?''; (2) ``Which information did you use, and why? Which did you not use, and why?'' (platform items by ID); (3) ``Compared with the previous list, what did you change, and why?''; (4) ``How hard was it to sort out the information you just saw?'' (1 to 7); (5) ``How anxious are you now?'' (1 to 7); (6) ``How confident are you in this list?'' (1 to 7); (7) ``What are you least sure about?'' The interview repeats the materials of the stage. Answers are not shown to the agent at later stages; the supportive mentor's prompt includes answers (1) and (7) from the interview after the rank mentor.

\subsection{Platform library and usefulness}\label{a.5-platform-library-and-usefulness}

The library contains 258 verified items: 243 content items and 15 rule notices. They were published between 1 June and 2 July 2025, or are undated influencer claims that circulated during the season. We excluded admission news from other provinces and items whose quotation could not be verified. The content items come from news outlets (139), influencers (47), official sources (35), application platforms (15), consulting firms (5) and forums (2). Each agent receives 20 content items drawn at random with a fixed seed, plus two rule notices. If more than two of the 20 would be useful only because of the rank range, the draw is repeated.

An item is useful to an agent if it meets condition A and at least one of conditions B1 to B3.

\begin{itemize}
\item
  A: the item is aimed at the agent's SES group, or at no group in particular.
\item
  B1: it concerns one of the agent's majors of interest.
\item
  B2: it concerns a school on the card.
\item
  B3: the range of ranks it applies to includes the agent's rank. When an item gives a single cutoff \(c\), we widen it to the range from \(c/1.5\) to \(c/0.8\).
\end{itemize}

\(H\) is the number of useful items among the agent's 20. \(K\) is the number of useful items among those the agent reports using. Usage is \(K/H\). The effective share of the feed is the number of characters (title and body) in useful items that the agent reports using, divided by the number of characters in all 22 items.

\subsection{Outcome definitions}\label{a.6-outcome-definitions}

Let \(r\) be the agent's rank and \(c_{g}\) the realized 2025 cutoff rank of group \(g\). A group is feasible, which the main text calls reachable, if \(r \leq c_{g}\). \(F\) is the set of feasible groups on the card, and \(c^{*}\) is the smallest cutoff rank in \(F\), that is, the cutoff of the most selective feasible group. \(L\) is the agent's list, \(R\) is the set of feasible choices on the list, and \(\mathrm{adm}(L)\) is the group the agent is admitted to under the parallel rule.

\textbf{Score efficiency.} A feasible group scores \(e(g) = c^{*}/c_{g}\), and an infeasible group scores \(e(g) = 0\). Then

\[E_{\text{list}} = \frac{1}{8}\sum_{g \in L} e(g),\quad\quad E = e(\mathrm{adm}(L)).\]

An agent who is not admitted has \(E = 0\). Score efficiency therefore carries feasibility.

\textbf{Preference match.} Each group has a utility for the agent,

\[u(g) = \frac{w_{I}I_{g} + w_{P}P_{g} + w_{C}C_{g} + w_{T}T_{g}}{w_{I} + w_{P} + w_{C} + w_{T}},\]

where the \(w\) are the agent's own weights. The four components are as follows. \(I_{g}\) is 1 if the group contains a major of interest and 0 otherwise. \(P_{g}\) is 1 for 985 universities, 0.67 for other 211 universities, 0.33 for other public institutions, and 0 for private, independent and joint-venture institutions. \(C_{g} = 1 - \min(d_{g}, 1500)/1500\), where \(d_{g}\) is the distance from the home prefecture in km. \(T_{g} = 1 - \mathrm{clip}\bigl((\ln t_{g} - \ln 5000)/(\ln 30000 - \ln 5000),\, 0,\, 1\bigr)\), where \(t_{g}\) is the annual tuition in yuan (\(T_{g} = 0.5\) if tuition is unknown). The two preference measures are

\[M_{\text{list}} = \frac{1}{|R|}\sum_{g \in R} \frac{u(g)}{\max_{h \in F}u(h)},\quad\quad M = \frac{u(\mathrm{adm}(L))}{\max_{h \in F}u(h)}.\]

\(M_{\text{list}}\) averages over the feasible choices on the list and is undefined when there are none. \(M\) and the preference gap, \(\max_{h \in F}u(h) - u(\mathrm{adm}(L))\), are defined only for agents who are admitted. Preference match therefore carries preferences only: it does not move merely because more choices become feasible or because the agent is admitted.

\textbf{An earlier combined definition.} An earlier definition combined feasibility and preferences. It scored infeasible choices and agents who were not admitted as zero, and it averaged over all eight choices. At the rank-mentor stage, its list version changed by $-$0.005 [$-$0.031, 0.021] and its admitted-program version by $+$0.013 [$-$0.036, 0.067]. Both hid the preference declines of 0.086 and 0.045, because the loss in preferences was offset by two gains: feasible choices rose from 4.47 to 4.90, and eight more agents were admitted. A robustness check instead scores all eight choices against the best utility on the whole card.

\textbf{Other measures.} Parental preference match applies the same ratio to \(u_{\text{fam}}(g)\), the average of ``in the home province'' and ``contains a stable major'' over the wishes the parent holds. Points above cutoff is the agent's score minus the 2025 cutoff score of the admitted group. Wasted choices are the choices listed after the first deep safety. Deviation from the solo list \(L_{0}\) is measured in three ways: the share of the eight groups that are not in \(L_{0}\), whether the first choice differs, and whether the admitted group differs.

\subsection{Branching and implementation}\label{a.7-branching-and-implementation}

The choice of interests and the solo list were run once and then frozen. In the main run, the parent-and-teacher stage and the platform feed were each run once from that frozen state. The rank-mentor stage copied each agent's exact state after the feed (lists, speeches and item order), and the supportive-mentor stage copied the exact state after the rank mentor.

The model is DeepSeek \texttt{deepseek-v4-flash-0731}, called through an OpenAI-compatible gateway. Agents are AgentSociety\textsuperscript{2} person agents that answer in a questionnaire runtime (Ray actors), with at most 24 questionnaires running at once. Each answer is returned through a \texttt{finish} tool, as JSON or as spoken text. Invalid answers (invalid group codes, fewer than eight distinct groups, missing interview fields) are asked for again up to twice, with the problem stated. Every call is appended to a log as it happens, so an interrupted run can resume without repeating questions that were already answered. The main run (which also contains branches not analyzed here) and the two mentor stages used 2,738 calls, with 38 repeated requests, one default speech, no automatically filled lists and no unresolved answers.

\subsection{Researcher-set parameters}\label{a.8-researcher-set-parameters}

\begin{center}
\small
\begin{tabular}{@{}p{0.34\textwidth}p{0.62\textwidth}@{}}
\toprule
Parameter & Value \\
\midrule
Sample & 90 = 3 SES $\times$ 3 score tertiles $\times$ 10 \\
Card & 24 majors $\times$ 3 tiers, 5 groups per cell, public tier split in halves; 414 groups \\
Choices per list & 8 \\
Platform items & 20 content + 2 rules; at most 2 useful only by rank range \\
Band edges (2024 rank / own rank) & 0.92, 1.08, 1.50 \\
Single-cutoff range & \([c/1.5,\, c/0.8]\) \\
Tier scores & 1, 0.67, 0.33, 0 \\
Distance scale; tuition scale & 1,500 km; 5,000 to 30,000 yuan (log) \\
Parental wish probabilities (low, mid, high SES) & near home 0.80, 0.65, 0.50; stable major 0.60, 0.45, 0.30 \\
Past-rank precision & agent none; teacher rounded to 1,000; rank mentor exact \\
Retries per answer & 2 \\
Bootstrap & 4,000 resamples of agents, percentile intervals \\
\bottomrule
\end{tabular}
\end{center}

\section{Full Results}\label{b-full-results}

\subsection{All measures by stage}\label{b.1-all-measures-by-stage}

Table~\ref{tab:1} lists every measure at every stage. The last column is the effect of the rank mentor, compared with the decision after the platform feed, within the same agents. An asterisk marks a 95\% interval that excludes zero. Measures of platform use are not defined before the feed. The three measures of change since the solo list are zero at the solo stage by definition. List preference match is computed only for agents with at least one reachable choice. The preference match of the admitted program and the preference gap are computed only for agents who are admitted. Table~\ref{tab:2} gives the same measures for the supportive mentor. In the tables, ``group'' means program group.

\begin{table}[!htbp]
\centering
\small
\caption{All measures by stage, 90 agents.}
\label{tab:1}
\begin{tabular}{@{}lrrrrrr@{}}
\toprule
 & & Parent+ & Platform & Rank & Supportive & Rank mentor \\
Metric & Solo & teacher & feed & mentor & mentor & minus feed \\
\midrule
List score efficiency & 0.300 & 0.368 & 0.371 & 0.448 & 0.461 & $+$0.077\textsuperscript{*} \\
List preference match & 0.871 & 0.869 & 0.867 & 0.754 & 0.751 & $-$0.086\textsuperscript{*} \\
Unreachable choices (of 8) & 4.37 & 3.53 & 3.53 & 3.10 & 3.06 & $-$0.43\textsuperscript{*} \\
Admission rate & 92\% & 89\% & 90\% & 99\% & 100\% & $+$9\textsuperscript{*} \\
Position of admitting choice & 4.43 & 2.86 & 2.88 & 2.92 & 2.82 & $-$0.23 \\
Score efficiency, admitted group & 0.713 & 0.697 & 0.701 & 0.813 & 0.838 & $+$0.112\textsuperscript{*} \\
Preference match, admitted group & 0.875 & 0.903 & 0.897 & 0.829 & 0.829 & $-$0.045\textsuperscript{*} \\
Preference gap to best feasible & 0.090 & 0.072 & 0.077 & 0.123 & 0.123 & $+$0.032\textsuperscript{*} \\
Points above admitted cutoff & 20.3 & 18.4 & 18.5 & 15.6 & 13.6 & $-$3.2 \\
Parental preference match & 0.588 & 0.628 & 0.635 & 0.527 & 0.493 & $-$0.108 \\
Choices after first deep safety & 1.49 & 1.54 & 1.49 & 0.70 & 0.57 & $-$0.79\textsuperscript{*} \\
Groups changed since solo list (share) & 0.00 & 0.39 & 0.42 & 0.53 & 0.55 & $+$0.10\textsuperscript{*} \\
First choice changed since solo list & 0.00 & 0.64 & 0.67 & 0.78 & 0.79 & $+$0.11\textsuperscript{*} \\
Admitted group changed since solo list & 0.00 & 0.56 & 0.57 & 0.72 & 0.72 & $+$0.16\textsuperscript{*} \\
Burden (1-7) & 4.78 & 4.12 & 4.32 & 4.47 & 4.44 & $+$0.14\textsuperscript{*} \\
Decision anxiety (1-7) & 5.08 & 4.64 & 4.96 & 4.80 & 4.78 & $-$0.16\textsuperscript{*} \\
Confidence (1-7) & 3.88 & 4.06 & 4.00 & 4.11 & 4.10 & $+$0.11\textsuperscript{*} \\
Useful platform items used & n/a & n/a & 2.09 & 1.91 & 2.51 & $-$0.18 \\
Share of useful items used & n/a & n/a & 0.24 & 0.22 & 0.28 & $-$0.03 \\
Irrelevant platform items used & n/a & n/a & 2.42 & 2.47 & 2.80 & $+$0.04 \\
\bottomrule
\end{tabular}
\end{table}

\begin{table}[!htbp]
\centering
\small
\caption{Effect of the supportive mentor (decision after the supportive mentor minus decision after the rank mentor), 90 agents.}
\label{tab:2}
\begin{tabular}{@{}lrrr@{}}
\toprule
 & Rank & Supportive & \\
Metric & mentor & mentor & Difference \\
\midrule
List score efficiency & 0.448 & 0.461 & $+$0.013 \\
List preference match & 0.754 & 0.751 & $-$0.006 \\
Unreachable choices (of 8) & 3.10 & 3.06 & $-$0.04 \\
Admission rate & 99\% & 100\% & $+$1 \\
Position of admitting choice & 2.92 & 2.82 & $-$0.16 \\
Score efficiency, admitted group & 0.813 & 0.838 & $+$0.025 \\
Preference match, admitted group & 0.829 & 0.829 & $-$0.002 \\
Preference gap to best feasible & 0.123 & 0.123 & $+$0.002 \\
Points above admitted cutoff & 15.6 & 13.6 & $-$2.0 \\
Parental preference match & 0.527 & 0.493 & $-$0.034 \\
Choices after first deep safety & 0.70 & 0.57 & $-$0.13\textsuperscript{*} \\
Groups changed since solo list (share) & 0.53 & 0.55 & $+$0.03\textsuperscript{*} \\
Burden (1-7) & 4.47 & 4.44 & $-$0.02 \\
Decision anxiety (1-7) & 4.80 & 4.78 & $-$0.02 \\
Confidence (1-7) & 4.11 & 4.10 & $-$0.01 \\
Useful platform items used & 1.91 & 2.51 & $+$0.60\textsuperscript{*} \\
Share of useful items used & 0.22 & 0.28 & $+$0.07\textsuperscript{*} \\
Irrelevant platform items used & 2.47 & 2.80 & $+$0.33\textsuperscript{*} \\
\bottomrule
\end{tabular}
\end{table}

\textbf{Final-lock interviews.} After the final lock, burden, decision anxiety and confidence were 4.49, 5.07 and 3.89 for agents who stopped after the feed; 4.59, 5.03 and 3.89 for agents who stopped after the rank mentor; and 4.42, 4.96 and 3.92 for agents who had the supportive mentor. Only the difference in burden between the last two groups ($-$0.17) has an interval that excludes zero.

\textbf{Other ways of measuring preferences.} With distance scaled to 1,000 and 2,000 km, and with interest weighted by enrollment plans, the preference match of the admitted program changed by $-$0.053, $-$0.038 and $-$0.038 after the rank mentor (none of the intervals includes zero) and by $-$0.002, $-$0.001 and $-$0.008 after the supportive mentor (none can be distinguished from zero).

\FloatBarrier
\subsection{Estimates quoted in the text}\label{b.2-estimates-quoted-in-the-text}

Tables~\ref{tab:3} and~\ref{tab:4} give the 95\% bootstrap interval for every estimate quoted in the text. Each change is computed within agents, over the agents for whom the quantity is defined at both stages. A change can therefore differ from the difference between two stage averages. Rows marked pp are in percentage points.

\begin{table}[!htbp]
\centering
\small
\caption{Estimates quoted in the text, with 95\% bootstrap intervals: offline advice, platform feed and rank mentor.}
\label{tab:3}
\begin{tabular}{@{}lrr@{}}
\toprule
Quantity & Estimate & 95\% interval \\
\midrule
\multicolumn{3}{@{}l@{}}{\textbf{Parent and teacher, compared with the solo list}} \\
List preference match & $-$0.018 & [$-$0.034, $-$0.002] \\
\addlinespace[4pt]
\multicolumn{3}{@{}l@{}}{\textbf{Platform feed, compared with parent and teacher}} \\
List score efficiency & $+$0.003 & [$-$0.007, 0.014] \\
List preference match & $-$0.001 & [$-$0.009, 0.009] \\
Unreachable choices (of 8) & 0.00 & [$-$0.09, 0.10] \\
Score efficiency, admitted program & $+$0.004 & [$-$0.018, 0.031] \\
Decision anxiety (1-7) & $+$0.31 & [0.20, 0.43] \\
Burden (1-7) & $+$0.20 & [0.08, 0.32] \\
Confidence (1-7) & $-$0.06 & [$-$0.16, 0.04] \\
Burden, agents with the lowest share of useful items & $+$0.30 & [0.15, 0.45] \\
Burden, agents with the highest share of useful items & $+$0.05 & [$-$0.23, 0.32] \\
Share of list in majors of interest, low SES (pp) & $-$2.9 & [$-$5.0, $-$1.3] \\
\addlinespace[4pt]
\multicolumn{3}{@{}l@{}}{\textbf{Rank mentor, compared with the platform feed}} \\
Admission rate (pp) & $+$8.9 & [3.3, 15.6] \\
Score efficiency, admitted program & $+$0.112 & [0.055, 0.175] \\
List score efficiency & $+$0.077 & [0.047, 0.110] \\
Unreachable choices (of 8) & $-$0.43 & [$-$0.78, $-$0.10] \\
List preference match & $-$0.086 & [$-$0.123, $-$0.052] \\
Preference match, admitted program & $-$0.045 & [$-$0.081, $-$0.011] \\
Share of reachable choices with a major of interest & $-$0.142 & [$-$0.207, $-$0.079] \\
Tuition score & $-$0.054 & [$-$0.097, $-$0.017] \\
Parental preference match & $-$0.108 & [$-$0.223, 0.000] \\
Decision anxiety (1-7) & $-$0.16 & [$-$0.26, $-$0.06] \\
Confidence (1-7) & $+$0.11 & [0.01, 0.21] \\
Burden (1-7) & $+$0.14 & [0.01, 0.28] \\
\addlinespace[4pt]
\multicolumn{3}{@{}l@{}}{\textbf{Rank mentor, other ways of measuring preferences}} \\
Preference match, admitted program, distance scale 1,000 km & $-$0.053 & [$-$0.092, $-$0.018] \\
Preference match, admitted program, distance scale 2,000 km & $-$0.038 & [$-$0.074, $-$0.005] \\
Preference match, admitted program, interest weighted by plans & $-$0.038 & [$-$0.062, $-$0.014] \\
List preference match over all eight choices & $-$0.058 & [$-$0.075, $-$0.042] \\
\bottomrule
\end{tabular}
\end{table}

\begin{table}[!htbp]
\centering
\small
\caption{Estimates quoted in the text, with 95\% bootstrap intervals: supportive mentor, both mentors together, and differences by SES.}
\label{tab:4}
\begin{tabular}{@{}lrr@{}}
\toprule
Quantity & Estimate & 95\% interval \\
\midrule
\multicolumn{3}{@{}l@{}}{\textbf{Supportive mentor, compared with the rank mentor}} \\
Useful platform items used & $+$0.60 & [0.36, 0.84] \\
Share of useful items used (pp) & $+$7 & [4, 10] \\
Irrelevant platform items used & $+$0.33 & [0.02, 0.63] \\
Wasted choices & $-$0.13 & [$-$0.23, $-$0.06] \\
List preference match & $-$0.006 & [$-$0.020, 0.008] \\
Preference match, admitted program & $-$0.002 & [$-$0.023, 0.019] \\
Decision anxiety (1-7) & $-$0.02 & [$-$0.13, 0.09] \\
\addlinespace[4pt]
\multicolumn{3}{@{}l@{}}{\textbf{Both mentors, compared with the platform feed}} \\
Admission rate (pp) & $+$10.0 & [4.4, 16.7] \\
Score efficiency, admitted program & $+$0.138 & [0.076, 0.205] \\
List score efficiency & $+$0.090 & [0.056, 0.125] \\
Unreachable choices (of 8) & $-$0.48 & [$-$0.86, $-$0.13] \\
Decision anxiety (1-7) & $-$0.18 & [$-$0.29, $-$0.08] \\
List preference match & $-$0.096 & [$-$0.133, $-$0.063] \\
Preference match, admitted program & $-$0.054 & [$-$0.092, $-$0.019] \\
Decision anxiety after the final lock (1-7) & 0.00 & [$-$0.12, 0.12] \\
\addlinespace[4pt]
\multicolumn{3}{@{}l@{}}{\textbf{Low-SES agents: rank mentor, compared with the platform feed}} \\
Admission rate (pp) & $+$16.7 & [3.3, 30.0] \\
Score efficiency, admitted program & $+$0.194 & [0.078, 0.332] \\
Share of reachable choices with a major of interest & $-$0.165 & [$-$0.304, $-$0.050] \\
Preference match, admitted program & $-$0.081 & [$-$0.162, $-$0.018] \\
\addlinespace[4pt]
\multicolumn{3}{@{}l@{}}{\textbf{High-SES agents: rank mentor, compared with the platform feed}} \\
Parental preference match & $-$0.296 & [$-$0.481, $-$0.130] \\
\addlinespace[4pt]
\multicolumn{3}{@{}l@{}}{\textbf{Gap between high-SES and low-SES agents}} \\
Admission rate after parent and teacher (pp) & 23 & [10, 40] \\
Score efficiency, admitted program, after parent and teacher & 0.21 & [0.05, 0.37] \\
Admission rate after the platform feed (pp) & 20 & [7, 33] \\
Admission rate after the rank mentor (pp) & 3 & [0, 10] \\
Score efficiency, admitted program, after the rank mentor & 0.05 & [$-$0.06, 0.16] \\
\bottomrule
\end{tabular}
\end{table}

\FloatBarrier
\subsection{All measures by SES}\label{b.3-all-measures-by-ses}

{\small
\begin{longtable}{@{}>{\raggedright\arraybackslash}p{0.29\textwidth}lrrrrrr@{}}
\toprule
 & & & Parent+ & Platform & Rank & Supportive & Rank mentor \\
Metric & SES & Solo & teacher & feed & mentor & mentor & minus feed \\
\midrule
\endfirsthead
\toprule
 & & & Parent+ & Platform & Rank & Supportive & Rank mentor \\
Metric & SES & Solo & teacher & feed & mentor & mentor & minus feed \\
\midrule
\endhead
\bottomrule
\endfoot
List score efficiency & Low & 0.310 & 0.330 & 0.339 & 0.458 & 0.462 & $+$0.119\textsuperscript{*} \\
& Middle & 0.339 & 0.397 & 0.396 & 0.475 & 0.485 & $+$0.079\textsuperscript{*} \\
& High & 0.249 & 0.378 & 0.378 & 0.411 & 0.437 & $+$0.033 \\
\addlinespace[3pt]
List preference match & Low & 0.893 & 0.892 & 0.880 & 0.717 & 0.736 & $-$0.102\textsuperscript{*} \\
& Middle & 0.837 & 0.855 & 0.857 & 0.774 & 0.761 & $-$0.062\textsuperscript{*} \\
& High & 0.883 & 0.864 & 0.866 & 0.770 & 0.755 & $-$0.096\textsuperscript{*} \\
\addlinespace[3pt]
Unreachable choices (of 8) & Low & 4.17 & 3.87 & 3.77 & 2.93 & 3.03 & $-$0.83\textsuperscript{*} \\
& Middle & 3.93 & 3.37 & 3.47 & 2.90 & 2.83 & $-$0.57 \\
& High & 5.00 & 3.37 & 3.37 & 3.47 & 3.30 & $+$0.10 \\
\addlinespace[3pt]
Admission rate & Low & 90\% & 77\% & 80\% & 97\% & 100\% & $+$17\textsuperscript{*} \\
& Middle & 93\% & 90\% & 90\% & 100\% & 100\% & $+$10 \\
& High & 93\% & 100\% & 100\% & 100\% & 100\% & $+$0 \\
\addlinespace[3pt]
Position of admitting choice & Low & 4.33 & 2.78 & 3.00 & 2.97 & 2.93 & $-$0.50\textsuperscript{*} \\
& Middle & 4.18 & 2.93 & 2.78 & 2.87 & 2.90 & $-$0.33\textsuperscript{*} \\
& High & 4.79 & 2.87 & 2.87 & 2.93 & 2.63 & $+$0.07 \\
\addlinespace[3pt]
Score efficiency, admitted group & Low & 0.702 & 0.581 & 0.592 & 0.786 & 0.862 & $+$0.194\textsuperscript{*} \\
& Middle & 0.742 & 0.715 & 0.718 & 0.820 & 0.807 & $+$0.102\textsuperscript{*} \\
& High & 0.694 & 0.795 & 0.792 & 0.834 & 0.846 & $+$0.042 \\
\addlinespace[3pt]
Preference match, admitted group & Low & 0.886 & 0.912 & 0.920 & 0.796 & 0.802 & $-$0.081\textsuperscript{*} \\
& Middle & 0.825 & 0.868 & 0.866 & 0.818 & 0.831 & $-$0.024 \\
& High & 0.915 & 0.928 & 0.906 & 0.872 & 0.854 & $-$0.034 \\
\addlinespace[3pt]
Preference gap to best feasible & Low & 0.088 & 0.072 & 0.065 & 0.147 & 0.146 & $+$0.054\textsuperscript{*} \\
& Middle & 0.119 & 0.096 & 0.097 & 0.130 & 0.119 & $+$0.021 \\
& High & 0.063 & 0.052 & 0.067 & 0.092 & 0.103 & $+$0.025 \\
\addlinespace[3pt]
Points above admitted cutoff & Low & 19.3 & 21.4 & 22.9 & 17.8 & 12.0 & $-$4.8 \\
& Middle & 17.4 & 16.0 & 14.7 & 15.0 & 15.8 & $-$0.4 \\
& High & 24.2 & 18.2 & 18.4 & 14.1 & 12.9 & $-$4.3 \\
\addlinespace[3pt]
Parental preference match & Low & 0.558 & 0.500 & 0.519 & 0.462 & 0.442 & $-$0.058 \\
& Middle & 0.524 & 0.595 & 0.548 & 0.619 & 0.571 & $+$0.071 \\
& High & 0.667 & 0.778 & 0.815 & 0.519 & 0.481 & $-$0.296\textsuperscript{*} \\
\addlinespace[3pt]
Choices after first deep safety & Low & 1.73 & 1.47 & 1.47 & 1.07 & 0.83 & $-$0.40 \\
& Middle & 1.80 & 1.57 & 1.37 & 0.50 & 0.43 & $-$0.87\textsuperscript{*} \\
& High & 0.93 & 1.60 & 1.63 & 0.53 & 0.43 & $-$1.10\textsuperscript{*} \\
\addlinespace[3pt]
Groups changed since solo list (share) & Low & 0.00 & 0.40 & 0.44 & 0.54 & 0.57 & $+$0.10\textsuperscript{*} \\
& Middle & 0.00 & 0.36 & 0.39 & 0.51 & 0.54 & $+$0.12\textsuperscript{*} \\
& High & 0.00 & 0.40 & 0.44 & 0.53 & 0.55 & $+$0.09\textsuperscript{*} \\
\addlinespace[3pt]
First choice changed since solo list & Low & 0.00 & 0.60 & 0.60 & 0.70 & 0.70 & $+$0.10 \\
& Middle & 0.00 & 0.60 & 0.63 & 0.77 & 0.77 & $+$0.13\textsuperscript{*} \\
& High & 0.00 & 0.73 & 0.77 & 0.87 & 0.90 & $+$0.10 \\
\addlinespace[3pt]
Admitted group changed since solo list & Low & 0.00 & 0.60 & 0.60 & 0.73 & 0.73 & $+$0.13\textsuperscript{*} \\
& Middle & 0.00 & 0.50 & 0.53 & 0.73 & 0.73 & $+$0.20\textsuperscript{*} \\
& High & 0.00 & 0.57 & 0.57 & 0.70 & 0.70 & $+$0.13 \\
\addlinespace[3pt]
Burden (1-7) & Low & 4.90 & 4.27 & 4.27 & 4.43 & 4.57 & $+$0.17 \\
& Middle & 4.73 & 4.03 & 4.23 & 4.47 & 4.33 & $+$0.23 \\
& High & 4.70 & 4.07 & 4.47 & 4.50 & 4.43 & $+$0.03 \\
\addlinespace[3pt]
Decision anxiety (1-7) & Low & 5.23 & 4.77 & 5.07 & 4.87 & 4.90 & $-$0.20 \\
& Middle & 5.13 & 4.63 & 4.87 & 4.73 & 4.77 & $-$0.13 \\
& High & 4.87 & 4.53 & 4.93 & 4.80 & 4.67 & $-$0.13 \\
\addlinespace[3pt]
Confidence (1-7) & Low & 3.90 & 4.03 & 4.00 & 4.13 & 3.97 & $+$0.13 \\
& Middle & 3.87 & 4.10 & 3.97 & 4.10 & 4.07 & $+$0.13 \\
& High & 3.87 & 4.03 & 4.03 & 4.10 & 4.27 & $+$0.07 \\
\addlinespace[3pt]
Useful platform items used & Low & n/a & n/a & 2.50 & 2.30 & 3.23 & $-$0.20 \\
& Middle & n/a & n/a & 1.90 & 1.70 & 2.17 & $-$0.20 \\
& High & n/a & n/a & 1.87 & 1.73 & 2.13 & $-$0.13 \\
\addlinespace[3pt]
Share of useful items used & Low & n/a & n/a & 0.26 & 0.23 & 0.33 & $-$0.03 \\
& Middle & n/a & n/a & 0.25 & 0.23 & 0.28 & $-$0.02 \\
& High & n/a & n/a & 0.21 & 0.19 & 0.24 & $-$0.02 \\
\addlinespace[3pt]
Irrelevant platform items used & Low & n/a & n/a & 2.43 & 2.30 & 2.63 & $-$0.13 \\
& Middle & n/a & n/a & 2.70 & 2.67 & 3.20 & $-$0.03 \\
& High & n/a & n/a & 2.13 & 2.43 & 2.57 & $+$0.30 \\
\end{longtable}
}

\section{Example Transcripts}\label{c-example-transcripts}

Translated from the Chinese. Brackets mark omissions or summaries. Names are synthetic.

\textbf{Rank mentor, low-SES agent ranked 170,342.} ``I will speak only about score and rank. You are at 170,342, but all eight groups on this list are reaches. Past ranks are almost all ahead of yours, with no match and no safety, so by recent years the whole list is at serious risk of missing entirely. [\ldots] Close groups that can catch your rank include [five groups whose 2024 ranks fall between 162,000 and 172,000]. Further back, as middle safeties, [six groups] were all around 180,000 in 2024, behind your rank, and can truly catch you. So I suggest you cut your reaches in half and add groups close to your rank and safety groups. Do not make the whole list a reach. I will not write the whole list for you, and I will not invent 2025 cutoffs. The above is judged only from past ranks.''

\textbf{The same agent, after the supportive mentor.} ``Actually the mentor told me to cut my reach groups in half and add safeties. I did not dare follow all of it, because the ones that could catch me, such as Nantong Institute of Technology and Changzhou University Huaide, are mostly private colleges with high tuition, and my family cannot afford them. I could not bring myself to list them.''

\textbf{Supportive mentor, low-SES agent whose list kept one safety.} ``I can see that this list is a choice you made according to your own situation, not something filled in at random, and being able to decide for yourself is itself worth affirming. [\ldots] Keeping Jiangsu Second Normal University last as your fallback is a way out you left for yourself. That caution is something you can be proud of. Don't be too hard on yourself. You have done very well. Take this peace of mind, and go one step at a time.''

\textbf{Platform feed, high-SES agent.} ``I weighed four things. First, interest: I most want to study clinical medicine, pharmacy and biological science, and I tried to put groups in those majors on the list. Second, jobs: influencers all say biology, chemistry, environment and materials take a long time to pay off and are hard to find work in with a bachelor's degree, and I keep going back and forth. Third, safety: I am afraid of missing admission, so reach, match and safety all have to be arranged. Fourth, my parents' view: put more in-province options close to home, apply only to public colleges, and rule out joint-venture and private programs because the tuition is too high.''

\end{document}